\documentclass[aps,prl,twocolumn,superscriptaddress]{revtex4}
\usepackage{graphicx}
\usepackage{color}
\usepackage{amsmath}
\usepackage{changes}

\begin{document}

\title{Paramagnetic Electronic Structure of CrSBr: Comparison between Ab Initio GW Theory and Angle-Resolved Photoemission Spectroscopy}

\author{Marco Bianchi}
\affiliation{Department of Physics and Astronomy, Interdisciplinary Nanoscience Center (iNANO), Aarhus University, 8000 Aarhus C, Denmark}
\author{Swagata Acharya}
\affiliation{National Renewable Energy Laboratories, Golden, CO 80401, USA}
\affiliation{Institute for Molecules and Materials, Radboud University, 6525 AJ Nijmegen, the Netherlands}
\author{Florian Dirnberger}
\affiliation{Institute of Applied Physics and W{\"u}rzburg-Dresden Cluster of Excellence ct.qmat, Technische Universit{\"a}t Dresden, Germany}
\author{Julian Klein}
\affiliation{Department of Materials Science and Engineering, Massachusetts Institute of Technology, Cambridge, Massachusetts 02139, USA}
\author{Dimitar Pashov}
\affiliation{King’s College London, Theory and Simulation of Condensed Matter, The Strand, WC2R 2LS London, UK}
\author{Kseniia Mosina}
\author{Zdenek Sofer}
\affiliation{Department of Inorganic Chemistry, University of Chemistry and Technology Prague, Technickaá 5, 166 28 Prague 6, Czech Republic}
\author{Alexander N. Rudenko}
\affiliation{Institute for Molecules and Materials, Radboud University, 6525 AJ Nijmegen, the Netherlands}
\author{Mikhail I. Katsnelson}
\affiliation{Institute for Molecules and Materials, Radboud University, 6525 AJ Nijmegen, the Netherlands}
\author{Mark van Schilfgaarde}
\affiliation{King’s College London, Theory and Simulation of Condensed Matter, The Strand, WC2R 2LS London, UK}
\affiliation{National Renewable Energy Laboratories, Golden, CO 80401, USA}
\author{Malte R\"osner}
\affiliation{Institute for Molecules and Materials, Radboud University, 6525 AJ Nijmegen, the Netherlands}
\author{Philip Hofmann}
\email{philip@phys.au.dk}
\affiliation{Department of Physics and Astronomy, Interdisciplinary Nanoscience Center (iNANO), Aarhus University, 8000 Aarhus C, Denmark}
\date{\today}

\begin{abstract}
  We explore the electronic structure of paramagnetic CrSBr by comparative first principles calculations and angle-resolved photoemission spectroscopy. We theoretically approximate the paramagnetic phase using a supercell hosting spin configurations with broken long-range order and applying quasiparticle self-consistent $GW$ theory, without and with the inclusion of excitonic vertex corrections to the screened Coulomb interaction (QS$GW$ and QS$G\hat{W}$, respectively). Comparing the quasi-particle band structure calculations to angle-resolved photoemission data collected at 200~K results in excellent agreement. This allows us to qualitatively explain the significant broadening of some bands as arising from the broken magnetic long-range order and/or electronic dispersion perpendicular to the quasi two-dimensional layers of the crystal structure. The experimental band gap at 200~K is found to be at least 1.51~eV at 200~K. At lower temperature, no photoemission data can be collected as a result of charging effects, pointing towards a significantly larger gap, which is consistent with the calculated band gap of $\approx$~2.1~eV.
\end{abstract}
\maketitle

Two-dimensional (2D) magnetic materials derived from van der Waals-bonded layered crystals have a wide range of potential applications due to the intricate coupling of spin, charge, and lattice degrees of freedom, and are fascinating in their own right, in particular since magnetic ordering in a 2D isotropic Heisenberg model is forbidden by the Mermin-Wagner theorem \cite{Mermin:1966uj}. Several examples of such materials have been discovered in the past few years, such as CrI$_3$ \cite{Huang:2017aa}, Cr$_2$Ge$_2$Te$_6$ \cite{Gong:2017aa} and Fe$_3$GeTe$_2$ \cite{Deng:2018aa}. Particularly promising properties are found in CrSBr, a layered magnetic semiconductor with an estimated band gap of 1.5~eV at room temperature \cite{Telford:2020tc,Klein:2022ua}. In CrSBr, the individual layers show ferromagnetic (FM) ordering that persists down to the monolayer limit with a high Curie temperature of approximately 150~K and an in-plane easy axis along the crystallographic $b$ direction \cite{Goser:1990aa,Telford:2020tc}. The alignment of magnetic moments between adjacent layers is antiferromagnetic (AFM) with a N\'{e}el temperature of 132~K.

Remarkably, charge transport \cite{Telford:2022uy,Wu:2022} and quasiparticle excitations such as excitons~\cite{Wilson:2021ty} or phonons \cite{Torres:2023,Pawbake.2023} are strongly correlated with the bulk magnetic order with a recent first promising demonstration of quantum transduction via exciton-magnon coupling~\cite{Bae:2022}. The magneto-correlated nature of CrSBr is complemented by a highly anisotropic crystal structure with strong implications on physical properties such as the dielectric screening, the exciton spectrum, quasiparticle interactions such as exciton-phonon and electron-phonon coupling \cite{Klein:2022ua} and the charge transport \cite{Wu:2022}. 

Fundamentally underlying these properties is the electronic band structure of CrSBr for which several predictions have been published, using different approximations \cite{Guo:2018wt,Jiang:2018ug,Wang:2019wf,Wang:2020wt,Wilson:2021ty,Klein:2022ua,Xu:2022uo,Klein:2022,Wu:2022}. All calculations do indeed reflect the expected anisotropy with a strongly anisotropic effective mass around the conduction band minimum and, to a much lesser degree, also in the valence band. However, existing calculations lack a description of the paramagnetic (PM) phase and an experimental determination of the band structure is also not reported to the best of our knowledge. Both are important ingredients to provide a clearer picture of the electronic properties and elucidate on the intricate interplay of magnetic order and electronic structure in CrSBr with the objective to understand its rich physical properties, quasiparticle excitations and interactions.

Here we explore the electronic band structure of CrSBr theoretically with a special emphasis on a consistent treatment of the non-trivial anisotrotropic screening and its effects to the Cr-$d$/S-$p$/Br-$p$ hybridization, and experimentally by angle-resolved photoemission spectroscopy (ARPES). We find that ARPES data cannot be acquired at low temperatures in the AFM ordered state because the sample becomes too insulating. The paramagnetic high temperature phase, on the other hand, is challenging to address by calculations because of the intrinsic spin disorder. However, as it turns out, theoretical aspects such as the role of the long-range Coulomb interaction have a much bigger effect on the band structure than the magnetic ordering such that a comparison between the AFM ground state calculation and the ARPES results remains meaningful. We obtain very good agreement between the paramagnetic ARPES data and an AFM calculation upon self-consistently including long-range Coulomb interactions and its screening. Our systematic comparison of supercell calculations within the layered FM and AFM states as well as in the paramagnetic phase together with a detailed analysis of the effects of the $k_z$ dispersion in these states finally allows us to uncover important broadening channels in the ARPES data.

\section{Theoretical and Experimental Details}

We apply three different levels of \textit{ab initio} theory: density functional theory (DFT) within the local-density approximation (LDA), quasiparticle self-consistent $GW$ theory (QS$GW$) \cite{qsgw}, which, in contrast to conventional $GW$ methods, modifies the charge density and is determined by a variational principle \cite{variational}, and QS$G\hat{W}$ \cite{Cunningham2023} in which the screened coulomb interaction $W$ is computed including vertex corrections (ladder diagrams) by solving a Bethe–Salpeter equation (BSE) within Tamm-Dancoff approximation \cite{hirata1999}. Crucially, our QS$G\hat{W}$ methods are fully self-consistent in both self-energy $\Sigma$ and the charge density \cite{acharya2021importance}. $G$, $\Sigma$, and $\hat{W}$ are updated iteratively until all of them converge. Our results are thus parameter-free and have no starting point bias. 

\begin{figure}
  \includegraphics[width=0.48\textwidth]{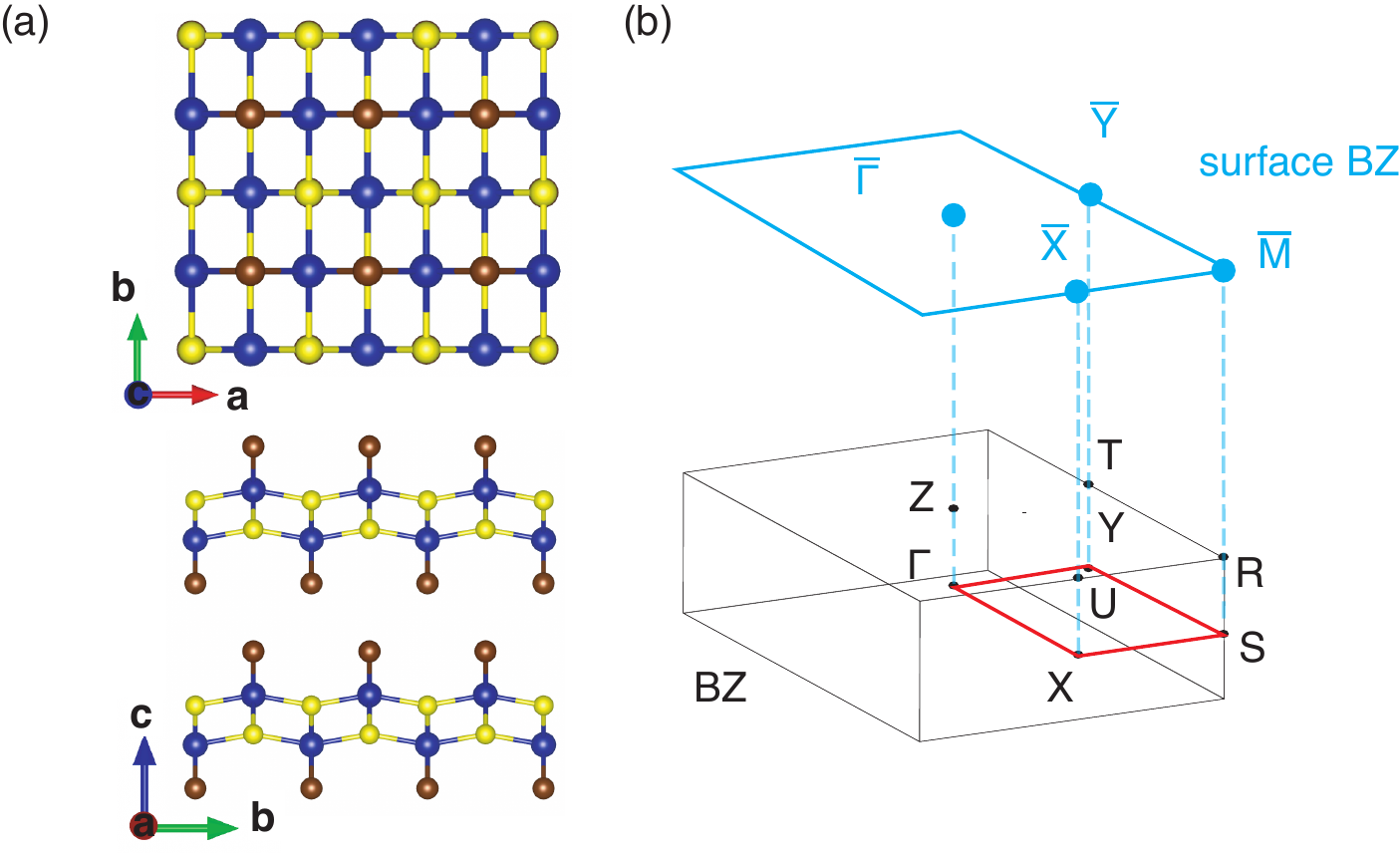}\\
  \caption{(a) Structure of the CrSBr unit cell. (b) Bulk (black) and (001) surface (cyan) Brillouin zones (BZs) for CrSBr with high symmetry points. Solid red lines the paths along which bands have been calculated.}
  \label{fig:structure}
\end{figure}

The CrSBr crystal structure is depicted in Fig.~\ref{fig:structure}(a). For all theoretical results we fix $a{=}3.504$, $b{=}4.738$, $c{=}2\times 7.907\,$\AA\ (the c-axis is doubled to treat antiferromagnetic order) to their experimental values~\cite{goser_magnetic_1990,lopez-paz_dynamic_2022} and relax all atomic positions in $z$ direction within VASP~\cite{Kresse1,Kresse2} on the GGA~\cite{pbe} level, using a collinear inter-layer antiferromagnetic (inter-layer ferromagnetic) ordering at a fixed inter-layer distance of $c/2$ and maintaining the orthorhombic crystal structure. This yields an intra-layer nearest neighbour Cr-Br distance of $2.50\,$\AA, as well as $2.37$ and $2.41\,$\AA\ for the two Cr-S distances, and an inter-layer nearest-neighbour Br-Br separation of $3.75\,$\AA\ in good agreement with Ref.~\onlinecite{lee_magnetic_2021}. All \textit{ab initio} band structure calculations are performed within Questaal using a full potential linear muffin tin orbital basis set. Single particle calculations (LDA, and energy band calculations with the static quasiparticlized QS$GW$ and  QS$G\hat{W}$ self-energy $\Sigma^{0}(k)$) are performed for both AFM and FM phases on a 12$\times$12$\times$6 \emph{k}-mesh while the (relatively smooth) dynamical self-energy $\Sigma(k)$ is constructed using a 6$\times$6$\times$3 \emph{k}-mesh. The QS\emph{GW} and QS$G\hat{W}$ cycles are iterated until the RMS change in $\Sigma^{0}$ reaches 10$^{-5}$\,Ry. 

For the magnetic configurations we considered layer polarized FM and AFM alignments with easy axes along the crystallographic $a$, $b$, and $c$  directions. We refer to these alignments by a notation combining the magnetic state and the easy axis, e.g., AFM-b. As we discuss in detail below, AFM-b and FM-b serve as good proxies for the paramagnetic phase, in which any long-range magnetic order has vanished, while short-range spin correlations persist. We validate this picture with QS$GW$ calculations in a paramagnetic $2 \times 2 \times 2$ supercell (48 atoms;  48  interstitial sites are added to augment the basis with floating orbitals). Local spin orientations are arranged in a quasirandom configuration to minimize the difference between the quasirandom and true random site correlation functions. An objective function composed from 480 pair and 384 triplet functions is minimized, following the approach by Zunger et al. \cite{zunger}.

CrSBr crystals were synthesized by chemical vapor phase growth \cite{Klein:2022}.  ARPES experiments were performed at the SGM-3 beamline of ASTRID2 \cite{Hoffmann:2004aa}. Energy and angular resolution were 60~meV and 0.2$^{\circ}$. The sample temperature was set to 200~K. The synchrotron radiation polarisation and the sample-to-analyzer direction were both in the plane of incidence and the analyzer slit was perpendicular to the plane of incidence. The CrSBr crystals were cleaved \emph{in situ} prior to measurements. The crystal orientation was determined by low energy electron diffraction and by inspection of the rectangular macroscopic crystal shape. 

\section{Results and Discussion}

\subsection{Self-Consistent $GW$ AFM results}

\begin{figure*}
  \includegraphics[width=1.0\textwidth]{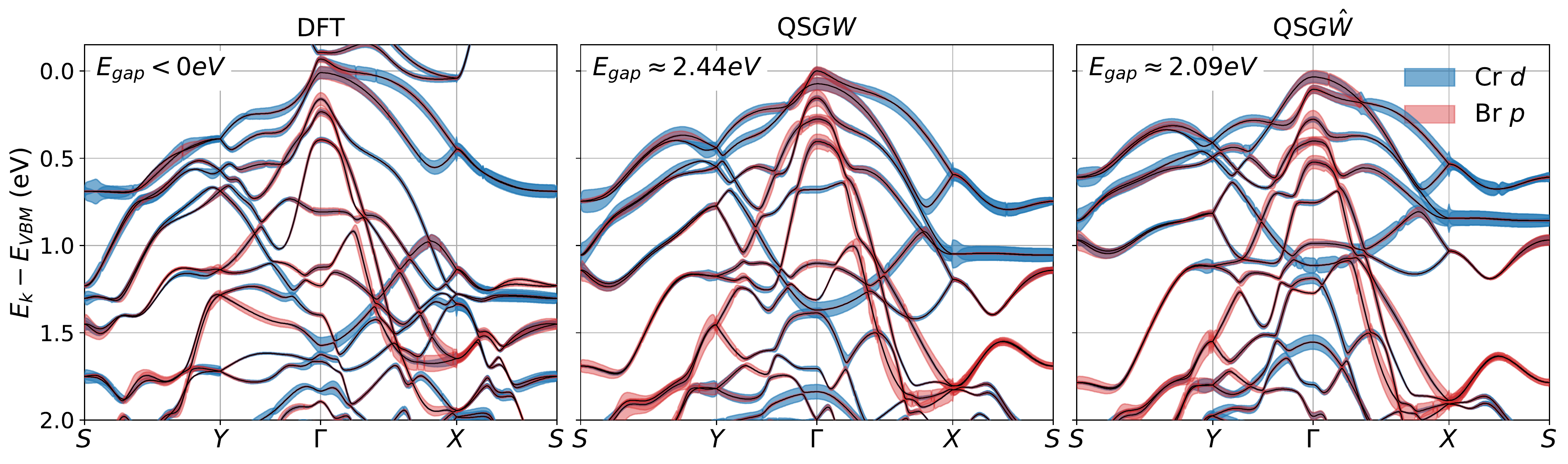}
  \caption{Orbital resolved quasi-particle band structures for AFM-b CrSBr within DFT, QS$GW$, and QS$G\hat{W}$. The individual (coloured) band widths indicate Cr $d$ orbital (blue) and Br $p$ orbital(red) contribution to each state. \label{fig:theoryMethods}}
\end{figure*}

We start by comparing band structure calculations for the AFM ground state with the easy axis in the $b$ crystallographic direction (AFM-b) within different levels of approximations to assess the role of non-local Coulomb interaction and screening together with modifications to the $d$-$p$ hybridization, as including these mechanisms turned out to be crucial for a correct description of CrX$_3$ compounds \cite{acharya2021electronic,acharya2022real}. Fig.~\ref{fig:structure} depicts the structure of CrSBr together with the bulk Brillouin zone (BZ). The projection of the bulk Brillouin zone along the crystallographic $c$ direction is also shown, as this is the relevant surface BZ for the cleaved crystals used in the ARPES experiment.
Fig.~\ref{fig:theoryMethods} summarizes the calculated bulk band structures obtained within DFT, QS$GW$, and QS$G\hat{W}$. Cr $d$ and Br $p$ orbital weights in each band are indicated in blue and red, respectively. 

The comparison between DFT and QS$GW$ shows striking differences as the DFT result does not show a finite band gap and as the valence band structures differ quantitatively and to large extent also qualitatively. The origin of this difference becomes clear upon inspecting the orbital contributions to the individual bands. From this we understand that the $d$-$p$ hybridization is strongly overestimated in DFT resulting in various overestimated avoided crossing gaps. 
Taking long-range Coulomb interaction renormalizations self-consistently into account  narrows the Cr $d$ dominated states, as visible at the $\Gamma$ point and reduces most of the avoided crossing gaps. 
As this is strongly governed by the $d$-$p$ hybridization, which is significantly affected by our $GW$ charge self-consistency, we stress that simplified DFT+$G_0W_0$ calculations, which neither update the charge density nor $W$ from changes in screening~\cite{Bruneval06b,Vidal10}, likely result in qualitatively wrong valence bands due to the inappropriate DFT starting point. The qualitative differences in the band structures around $\Gamma$, $S$, and $Y$ induced by self-consistency are in good agreement with our ARPES data, as we show in more detail below. 

Further corrections to the QS$GW$ results in form of excitonic screening to the Coulomb interaction, as rendered in QS$G\hat{W}$ calculations, do not qualitatively change the band structure anymore, but are responsible for quantitative differences mostly notable in all states with significant Cr $d$ character.

\subsection{Impact of Magnetic Disorder and Paramagnetic Properties}

\begin{figure*}
  \includegraphics[width=0.66\textwidth]{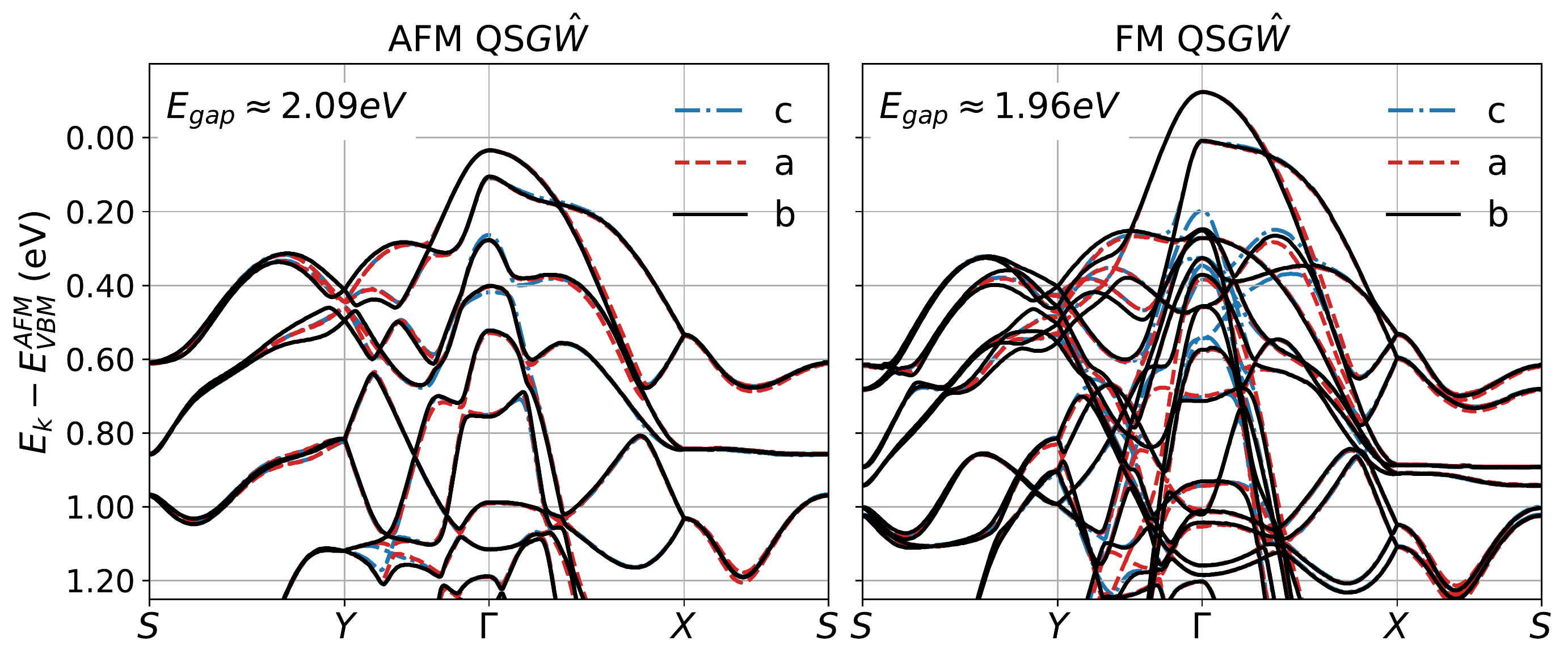}
  \caption{QS$G\hat{W}$ band structures for different magnetic easy axes. 
  \label{fig:theoryBSEABC}}
\end{figure*}

A comparison to the ARPES data taken in the high temperature paramagnetic phase should be undertaken using calculations representing the same phase. This is, however, highly challenging for theory due to the complexity resulting from the vanished long-range and residual short-range magnetic order in the paramagnetic phase. Thus, before we discuss our approximate paramagnetic results obtained from supercell calculations, we first compare QS$G\hat{W}$ calculations for AFM and FM configurations with varying spin easy axes in Fig.~\ref{fig:theoryBSEABC} as possible proxies for the PM phase. On the level of the quasi-particle energies, we do not find qualitative differences between the various easy axes reflecting the vanishing impact of magneto-crystalline anisotropy in the ordered phases. There are, however, a few quantitative details, which vary upon rotating the easy axis from AFM-b to AFM-a or AFM-c. This is accompanied by changes to the relative Cr $d$ weights. The density of states is nearly unaffected by the easy axis rotations and the band gap is not modified at all. 

As the AF magnetic exchange interaction between the layers is significantly weaker than the FM in-plane interactions, we expect that the inter-layer AFM long-range order is first broken as the temperature increases slightly above the Curie temperature, such that the  the paramagnetic phase may be approximately seen as a disordered inter-layered AFM/FM  structure with long-range in-plane FM order. The effect of such a scenario is best tested by comparing QS$G\hat{W}$ AFM and FM calculations, assuming that the experimental situation lies in between these extremes. The corresponding FM results are shown in the right panel of Fig.~\ref{fig:theoryBSEABC}. Easy axis rotation has only minor effects, also in the FM ordered phase. However, changing the magnetic polarization between the layers from AFM to FM has a significant impact on the band structure and the density of states. Due to the broken (magnetic) inversion symmetry in the FM phase, all degeneracies of the AFM band structure are lifted, which is most prominently seen in the split highest VBs in the FM case. Upon aligning the AFM and FM band structures with respect to their Fermi level differences (ca. 125\,meV), we find a major modification to the two upmost FM VBs, which leak into the AFM band gap and thereby reduce the total gap (as the FM CB position is nearly unaffected, see Appendix). All other states experience only quantitative renormalizations (next to the broken degeneracies).
When comparing the calculations to the experiment, magnetic disorder along the lines calculated here can be expected to result in a significant broadening of the bands. The effect will be especially pronounced in the two bands near the VBM. 

\begin{figure*}
  \includegraphics[width=0.99\textwidth]{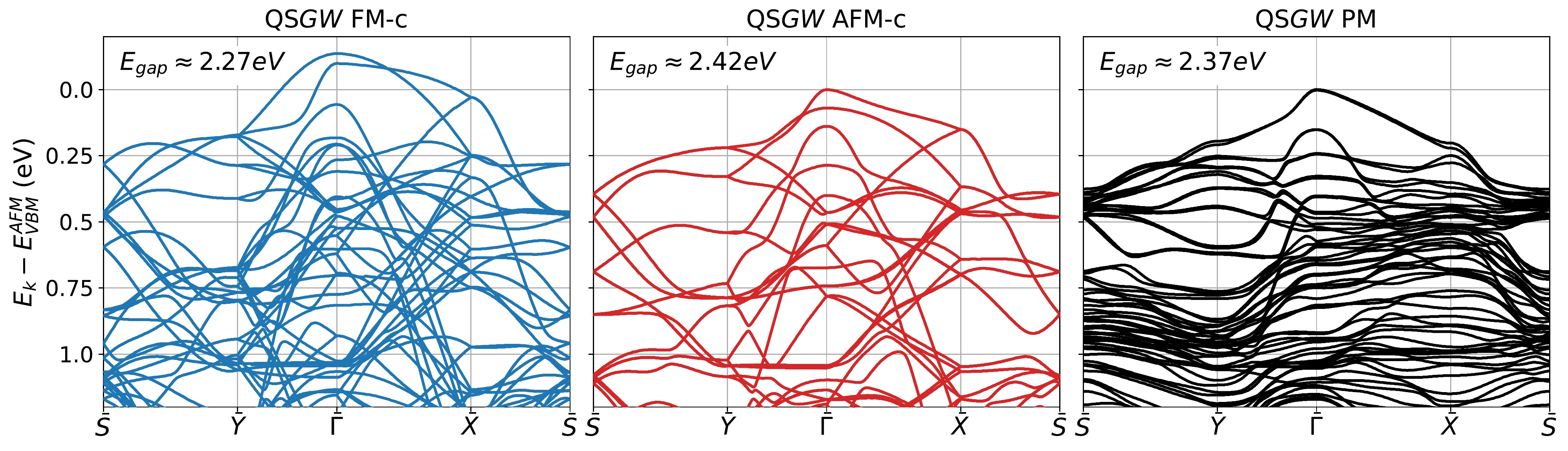}
  \caption{QS$GW$ band structures for supercells in FM-c and AFM-c together with the paramagnetic approximation. \label{fig:theorySupercell}}
\end{figure*}

This picture is verified and augmented by our paramagnetic $2\times2\times2$ supercell calculation within QS$GW$ approximation using frozen spins and including spin-orbit coupling shown in Fig.~\ref{fig:theorySupercell} together with the ordered FM-c and AFM-c band structures for the same supercells. Please note that the indicated high-symmetry points here refer to the ones in the reduced Brillouin zone of the supercell. The resulting band structures are thus folded with respect to those discussed before. The comparison between the the FM-c and AFM-c band structures resembles the one discussed above. The PM calculation shows many more finely split bands. This is a consequence of the spin disorder, which effectively broadens the states and which can be understood as a static version of spin fluctuations. Throughout the depicted band structures, we find many qualitatively similarities between the PM results and both, the FM and AFM ones. The most striking features of the PM valence bands are the significant dispersion in $k_z$ between $\Gamma$ and $Z$, which is also present in the FM data, but nearly completely suppressed in the AFM band structure (see Fig.~\ref{fig:theoryGWSupKz} in the Appendix), and the strong qualitative difference between the upmost valence bands. In the PM case these bands neither clearly resemble the FM or AFM ones, emphasizing that renormalization is a strong signature of the PM phase.

\subsection{ARPES Results}

\begin{figure}
 \includegraphics[width=0.5\textwidth]{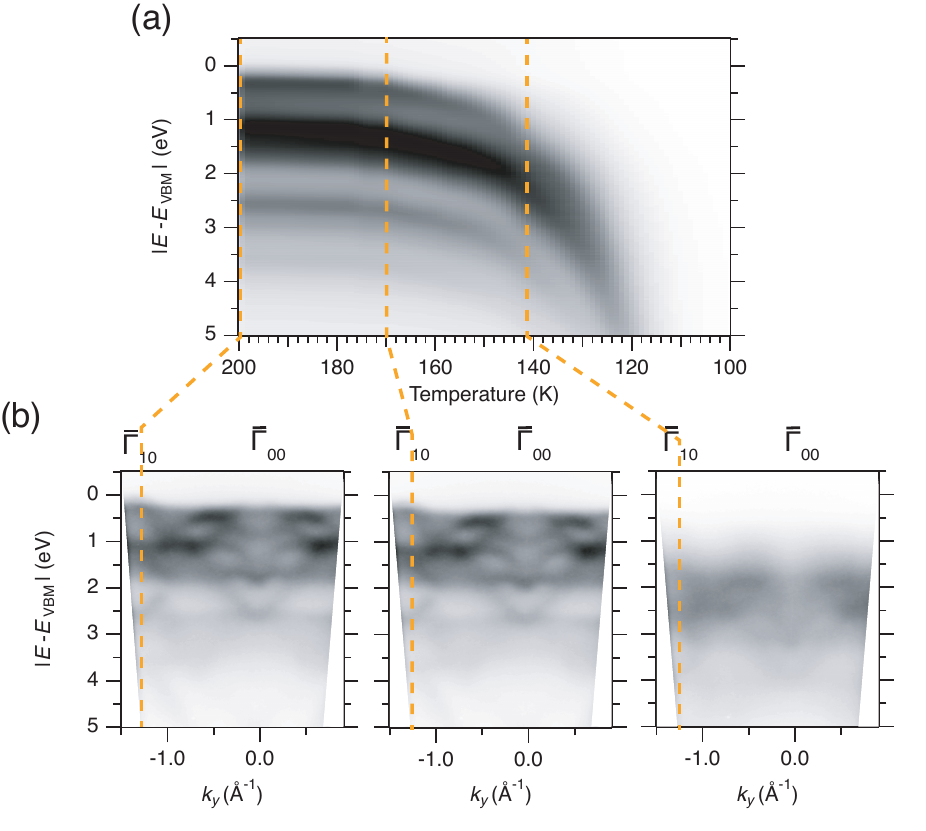}\\
  \caption{Illustration of charging at low temperature. (a) Photoemission intensity as a function of energy and temperature at $\overline{\Gamma}_{10}$ for h$\nu$=100~eV (for the definition of the high symmetry points see Fig.\ref{fig:100ev_exp}). (b) Photoemission intensity at $k_x$=0~\AA$^{-1}$ for three selected temperatures. Strong charging leads to an energy shift of the spectra below T$\approx$170~K.}
  \label{fig:charging}
\end{figure}

When collecting temperature-dependent ARPES spectra, charging effects were observed below $T{\approx}160$\,K. These were sufficiently strong to prevent measurements at low temperatures and the data reported here were therefore collected at 200~K, i.e., in the PM phase. The effect is illustrated in Fig.~\ref{fig:charging}  by a series of ARPES spectra taken while the sample was cooled down from T=200~K to 100~K.  The transition temperature to the insulating state of about 170~K is close to magnetic ordering temperature  \cite{Lee:2021to}. However, the effect is still surprising since the magnetic transition has only a small effect on the conductance determined in bulk transport measurements \cite{Telford:2020tc}. Also, depending on the chosen approximation, the magnetic order does affect the calculated band gap (see labels in Figs.~\ref{fig:theoryBSEABC} and \ref{fig:theorySupercell} and Appendix) but the differences are quite small.  

\begin{figure}
  \includegraphics[width=0.5\textwidth]{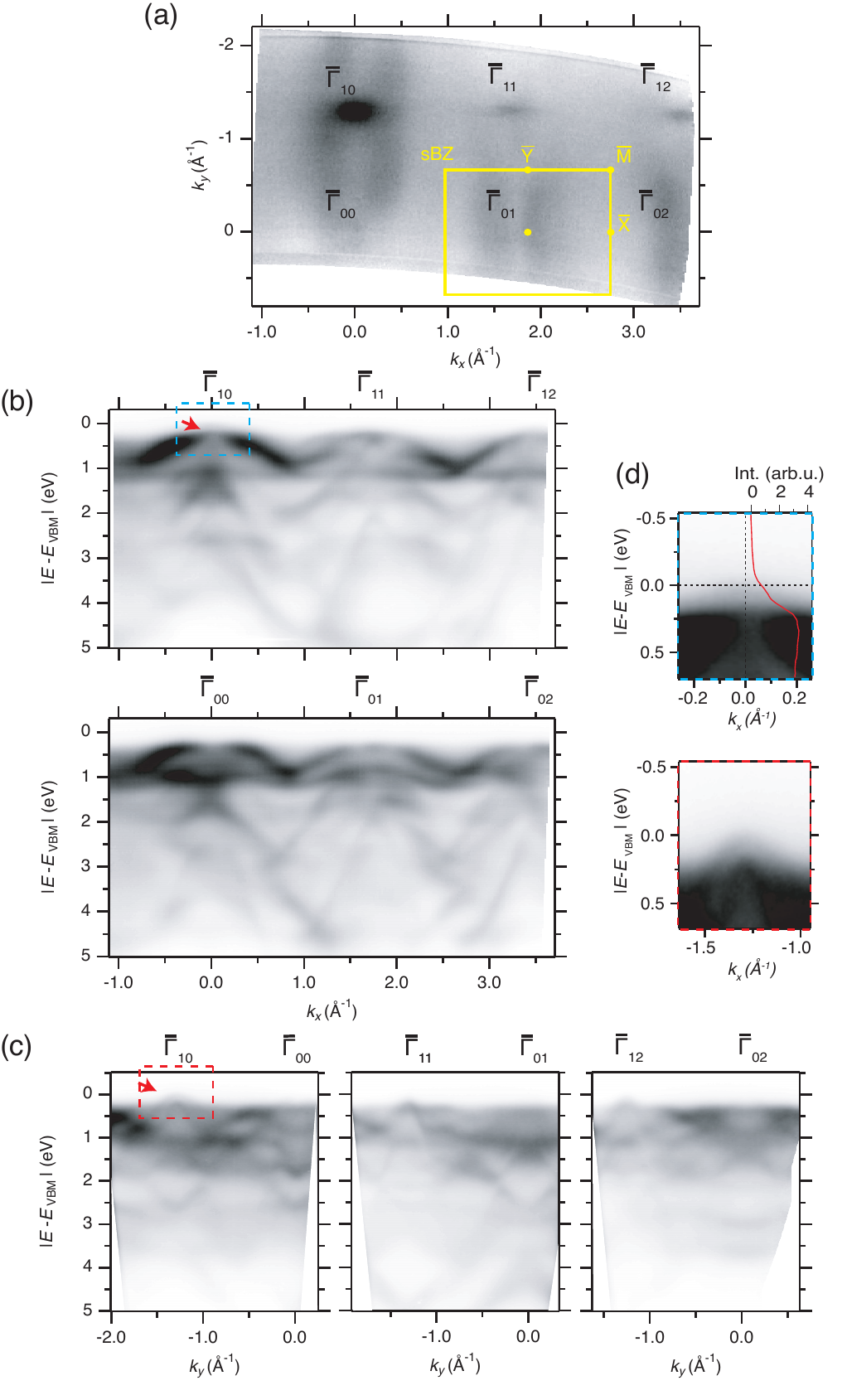}\\
  \caption{Electronic structure of CrSBr determined by ARPES at a photon energy of h$\nu$=100~eV. (a) Photoemission intensity integrated in a 30~meV window around the VBM. (b) Photoemission intensity in the $k_x$ direction along $\overline{\Gamma}_{10}-\overline{\Gamma}_{11}-\overline{\Gamma}_{12}$ and $\overline{\Gamma}_{00}-\overline{\Gamma}_{01}-\overline{\Gamma}_{02}$. (c) Corresponding cuts in the $k_y$ direction along the $\overline{\Gamma}_{10}-\overline{\Gamma}_{00}$, $\overline{\Gamma}_{11}-\overline{\Gamma}_{01}$ and $\overline{\Gamma}_{12}-\overline{\Gamma}_{02}$ directions. The red arrows mark the point at which the VBM is observed. (d) Magnification of the VBM region in panels (b) and (c). The red curve in the upper panel is the photoemission intensity at $\overline{\Gamma}_{10}$ as a function of energy. 
  }
  \label{fig:100ev_exp}
\end{figure}

Figure \ref{fig:100ev_exp} provides an overview of the ARPES results collected in an extended zone scheme. The photoemission intensity is shown as cuts through a three-dimensional data set, collected as a function of the energy and a 2D $\mathbf{k}=(k_x,k_y)$ parallel to the surface. The energy zero has been set to the estimated position of the VBM as observed around the  $\overline{\Gamma}_{10}$ point (the reason for choosing this particular point will become clear later). The Fermi energy of a metal in contact with the sample is at $E_\mathrm{F}=-1.51\,\mathrm{eV}$. Since no indications of the conduction band are found in ARPES, this implies that the band gap is at least 1.5~eV wide, consistent with the 1.5~eV gap from scanning tunnelling spectroscopy at room temperature \cite{Telford:2020tc,Klein:2022ua}. However, a gap of around 1.5~eV would imply a very strong, essentially degenerate, $n$ doping of the sample and one should not expect to observe the aforementioned charging effects, especially if the magnetic transition has only a minor effect on the gap. We thus take this as a strong indication that the band gap of CrSBr is  significantly larger. A gap size of $\approx$~2.0~eV as suggested by the calculations in Fig. \ref{fig:theoryBSEABC} could, in combination with high crystalline quality, explain the observed charging effects. In this case, the situation would be similar to that in high-quality organic crystals where charging is a well-known issue for ARPES investigations \cite{Nakayama2020}.

In Figure \ref{fig:100ev_exp}(a) the photoemission intensity is shown as a function of the 2D $\mathbf{k}$, integrated over an energy range of 30~meV around to the VBM. The $\mathbf{k}$ scan range is wide enough to enclose several surface BZs. The periodicity along $k_x$ (in the real space $a$ direction) is easily identified by the repeated pattern of high intensity points near $\overline{\Gamma}_{10}$, $\overline{\Gamma}_{11}$ and $\overline{\Gamma}_{11}$. It agrees with the expected length of a reciprocal lattice vector in this direction of 1.79~\AA$^{-1}$. The periodicity in the $k_y$ direction, in contrast, is not immediately obvious. The size of the BZ in the $y$ direction is 1.33~\AA$^{-1}$ but the photoemission intensity at the nominally equivalent $\overline{\Gamma}$ points does not appear to reflect this. In particular, we find local maxima in the photoemission intensity at  $\overline{\Gamma}_{10}$, $\overline{\Gamma}_{11}$ and $\overline{\Gamma}_{12}$ but local minima at  $\overline{\Gamma}_{00}$ (corresponding to normal emission) and $\overline{\Gamma}_{01}$.

The periodicity is clarified when inspecting cuts through the data set along high symmetry directions. Cuts in the $k_x$ direction along the  $\overline{\Gamma}_{10}-\overline{\Gamma}_{11}-\overline{\Gamma}_{12}$ and $\overline{\Gamma}_{00}-\overline{\Gamma}_{01}-\overline{\Gamma}_{02}$ lines are given in Fig. \ref{fig:100ev_exp}(b). Following the dispersions, it is clear that the assigned $\overline{\Gamma}$ and $\overline{X}$ are indeed high symmetry points and that the VBM is found at  $\overline{\Gamma}$. However, the apparent dispersion around the  $\overline{\Gamma}$ points is quite different. Around $\overline{\Gamma}_{10}$ the intensity of the bands at the highest energy appears to collapse but a weak intensity disperses to a higher energy than at any other $\overline{\Gamma}$ point (marked by a red arrow, see also Fig. \ref{fig:100ev_exp}(c) and (d) for a magnification of this region). This is consistent with Fig. \ref{fig:100ev_exp}(a) where the highest overall intensity is observed at $\overline{\Gamma}_{10}$ and this maximum in the dispersion is taken to be the VBM (defined as the position of the photoemission intensity's leading edge in in Fig. \ref{fig:100ev_exp}(d)). Near $\overline{\Gamma}_{11}$ and $\overline{\Gamma}_{12}$ it is easier to see a local maximum in the dispersion but this is found at a slightly lower energy than at $\overline{\Gamma}_{10}$. In fact, the same band is also visible at  $\overline{\Gamma}_{10}$, $\approx$ 170~meV below the very faint band forming the VBM. This is best seen in the magnification of the dispersion around the VBM at $\overline{\Gamma}_{10}$ shown in Fig. \ref{fig:100ev_exp}(d).
An extreme case is normal emission, $\overline{\Gamma}_{00}$, where the photoemission intensity around the VBM appears to be strongly suppressed. A similar picture presents itself at lower energy where the observable bands around equivalent high symmetry points do not appear to be the same with some exceptions, e.g., the band crossing at the $\overline{X}$ points at about 2~eV that is clearly observed  between  $\overline{\Gamma}_{10}$ and $\overline{\Gamma}_{11}$ and between $\overline{\Gamma}_{11}$ and $\overline{\Gamma}_{12}$.

The dispersion in the $k_y$ direction shows similar effects. Again, the local symmetry around the $\overline{\Gamma}$ and $\overline{Y}$ points in Fig. \ref{fig:100ev_exp}(c) is clear, confirming the expected periodicity. But also here, the local band structure around the symmetry points appears different for symmetry-equivalent points apart from some bands, for example a pronounced feature at 2.6~eV that is observed around $\overline{\Gamma}_{00}$, $\overline{\Gamma}_{01}$ and $\overline{\Gamma}_{10}$. 

The apparently different dispersion around symmetry-equivalent points could have several reasons. The first is that scanning the parallel crystal momentum $\mathbf{k}$ across the surface also leads to a change of the probed $k_z$. The different high symmetry points thus have the same $\mathbf{k}$ modulo a surface reciprocal lattice vector but they do not have the same $k_z$. Any dispersion along $k_z$ would thus lead to a different observed band structure around equivalent high symmetry points in the extended surface BZ. While this explanation might play a role, the effect is likely to be  minor because a pronounced $k_z$ dispersion is not expected for a layered van der Waals material such as CrSBr. Moreover, very similar band dispersions are made when using different photon energies, also supporting the 2D character of the band structure (see Appendix).
On the other hand, both calculations and ARPES results show a small degree of $k_z$ dispersion for the lower lying bands which could contribute to a broadening, as discussed below. 

A more likely explanation for the observed differences in apparent dispersion are strong matrix element effects which completely suppress some bands in certain regions of reciprocal space. A particularly interesting example is the VBM which is most clearly observed around the  $\overline{\Gamma}_{10}$ point and not at all around normal emission at  $\overline{\Gamma}_{00}$. Such pronounced matrix element variations are most likely explained by sub-lattice interference effects due to the presence of two equivalent atoms (of each kind) in the unit cell, similar to what is seen in the $\pi$, $\sigma$ and core level states of graphene \cite{Shirley:1995aa,Lizzit:2010aa}, and in particular near the top of the graphite and graphene $\sigma$ band that shows an intensity collapse exactly at normal emission, but not for the $\Gamma$ points in the neighbouring BZs \cite{Shirley:1995aa,Mazzola:2013aa,Mazzola:2017aa}. In the case of CrSBr, the situation is more complex because of the change of orbital character very near to the VBM (see Fig.~\ref{fig:theoryMethods}). In any event, the strong matrix element effects call for some care when comparing the ARPES results to calculations because the absence of dispersive features in the experiment does not necessarily indicate poor agreement with a calculation -- the features might just be suppressed by matrix element effects. In fact, the predicted state might be observable along an equivalent path in another BZ. 

\begin{figure}
 \includegraphics[width=0.5\textwidth]{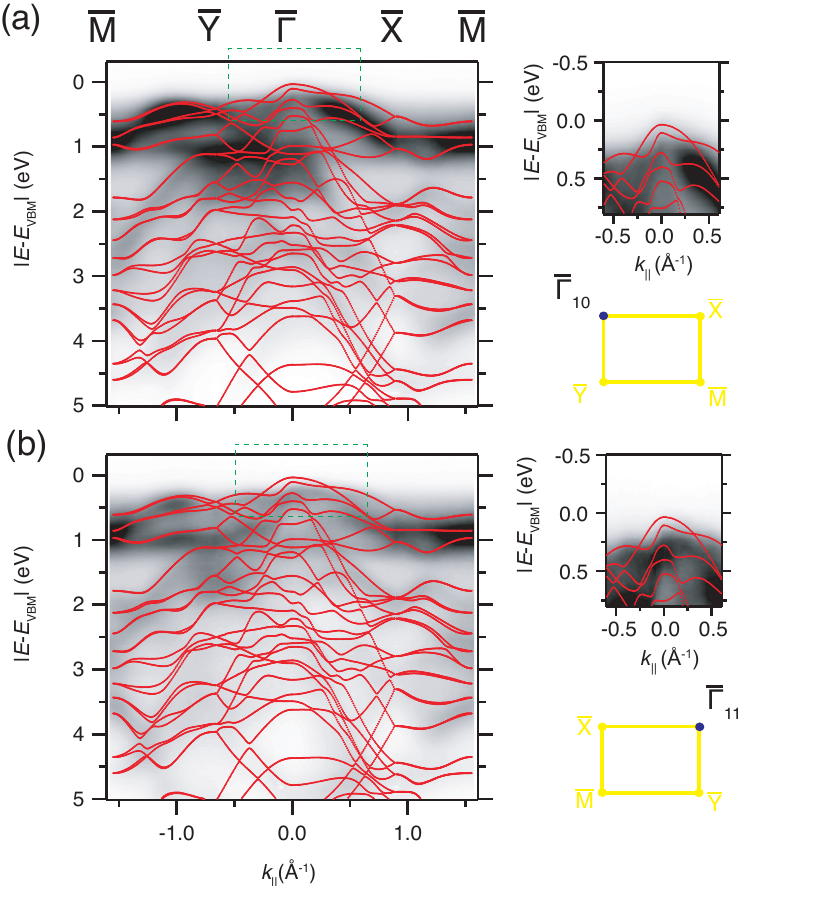}\\
  \caption{Photoemission intensity and AFM-b QS$G\hat{W}$ calculation (h$\nu$=100~eV) for two equivalent but different paths in the extended zone scheme (indicated on the right hand side).  
  }
  \label{fig:100ev_expthe}
\end{figure}

Keeping these considerations in mind, we proceed to a comparison between the measured and calculated band structures. To this end, we use the  AFM-b QS$G\hat{W}$ calculation. Fig. \ref{fig:100ev_expthe} shows this comparison for experimental data along two different but equivalent paths in the extended zone scheme, one including the $\overline{\Gamma}_{10}$ point and one including the $\overline{\Gamma}_{11}$ point. The energy zero in the experimental bands is still the estimated position of the VBM as observed around $\overline{\Gamma}_{10}$.  The experimental bands are quite broad so that structures often contain more than one calculated band. The overall agreement is, however, excellent and we find QS$G\hat{W}$ states at all locations with increased ARPES intensity, as well as clearly vanishing ARPES intensity in regions with no QS$G\hat{W}$ states. We stress that this is so for the entire wide energy range studied, not only near the VBM. This would clearly not be the case for a comparison to the metallic DFT band structure and the agreement thus strongly supports the necessity to go beyond DFT for this material.

The situation around the VBM deserves special attention. As already seen in Fig. \ref{fig:100ev_exp}, the highest energy band is observed at  $\overline{\Gamma}_{10}$ and this is assigned to the VBM. The band is broad and has a very low intensity. Indeed, it is only clearly visible in the magnification of Fig.~\ref{fig:100ev_exp}(d). Keeping in mind the discussion accompanying Figs. \ref{fig:theoryBSEABC} and \ref{fig:theorySupercell}, this observation is consistent with the expectation of a strongly broadened VBM in the presence of magnetic disorder in the paramagnetic state. The band is too broad and weak for a quantitative determination of the dispersion and it is thus not possible to verify the predicted anisotropy of the effective mass.
The local band maxima at $\overline{\Gamma}_{11}$ and $\overline{\Gamma}_{12}$ appear at a slightly lower energy ($\approx$ 170~meV) and are therefore assigned to the lower lying maximum of pure Cr $d$ character. This also appears to match very well with the superimposed calculations. 

\begin{figure}
 \includegraphics[width=0.5\textwidth]{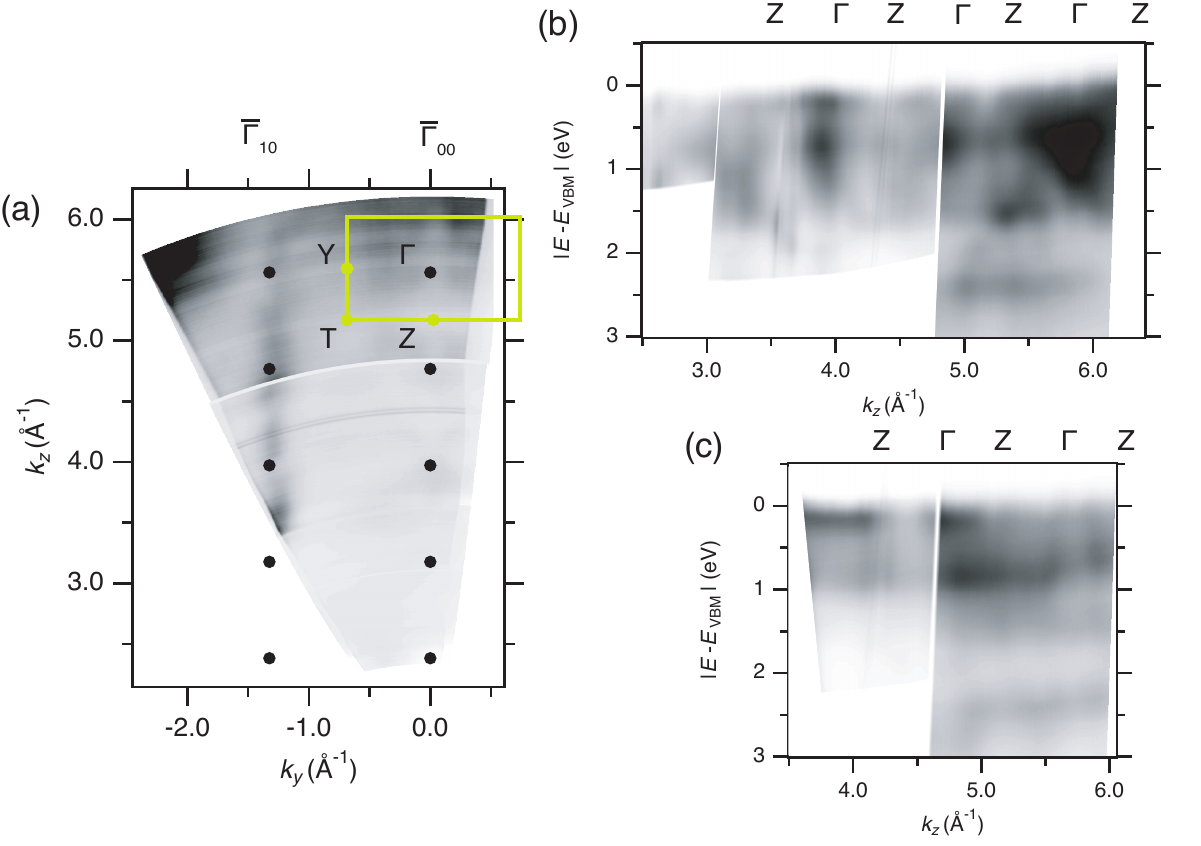}\\
  \caption{Dispersion along $k_z$ (a) Constant energy surface in a range of 30~meV around VBM as a function of  $k_z$ and $k_y$.  Consecutive $\Gamma$ points are marked in blue while in green the BZ. b) and c) energy dispersion as a function of $k_z$ at $k_y=0$~\AA$^{-1}$ (i.e., $\overline{\Gamma}_{00}$) and $k_y=1.33$~\AA$^{-1}$ (i.e., $\overline{\Gamma}_{10}$), respectively. 
  }
  \label{fig:kz}
\end{figure}

Finally, we experimentally explore the dimensionality of the electronic structure in CrSBr by a photon energy scan to systematically vary $k_z$.  Fig. \ref{fig:kz}(a) shows the photoemission intensity integrated in an energy window of 30~meV around the VBM as a function of $k_y$ and $k_z$. A cut through the bulk BZ is superimposed and the bulk $\Gamma$ points are indicated by blue markers. The $k_z$ values have been obtained from the photon energy under the assumption of free electron final states, using an inner potential of 12.8~eV. The vertical streak of photoemission intensity around $\overline{\Gamma}_{10}$ indicates a two-dimensional electronic structure around the VBM. The situation around $\overline{\Gamma}_{00}$ is inconclusive because the VBM is not observed there (see Fig. \ref{fig:100ev_exp}).   Fig. \ref{fig:kz}(b) and (c) show the photoemission intensity as a function of energy and $k_z$ for $k_y$ fixed to $\overline{\Gamma}_{00}$ and $\overline{\Gamma}_{10}$, respectively. There is very little evidence of dispersion in the $k_z$ direction here as well. The most intense features do not show any $k_z$ dependence of their binding energy and can be easily identified with the most intense features seen for a photon energy of 100~eV ($k_z\approx$5.29~\AA$^{-1}$) in Fig. \ref{fig:100ev_exp}(c). The lack of $k_z$ dispersion is also confirmed by a data set showing a very similar $\mathbf{k}$ dispersion throughout the extended zone, obtained using a different photon energy and shown in the Appendix. On the other hand, some degree of $k_z$ dispersion appears to be present at higher lying bands. In Fig.~\ref{fig:kz}(b), for instance, a dispersing state is visible around 1~eV between  4~\AA$^{-1}$ and 5~\AA$^{-1}$ and another one around 2~eV between 5~\AA$^{-1}$ and 5.7~\AA$^{-1}$. Such a dispersion in the lower lying bands is consistent with the calculated $k_z$-dependent band structures in paramagnetic phase in Fig.~\ref{fig:theoryGWSupKz} and this $k_z$ can thus be expected to contribute to the energy broadening of the lower lying bands. 

\section{Conclusions}

When describing the electronic structure of CrSBr theoretically, it is essential to include the long-range Coulomb interaction and its impact on the charge density self-consistently, as it is done within the QS$GW$ and QS$G\hat{W}$ approximations. The quasi-particle band structures calculated in this way are in excellent agreement with ARPES results. DFT as well as non-self-consistent DFT+$G_0W_0$ calculations, on the other hand, result in band structures that are significantly different from our experimental data. 

Both ARPES measurements and QS$G\hat{W}$ calculations show a small degree of dispersion along $k_z$ for states close to the VBM and we find that AFM-b and FM-b configurations are good proxies for describing the paramagnetic phase. The comparison of these calculations suggests that magnetic disorder in the paramagnetic phase would have a particularly strong effect on the bands forming the VBM, giving rise to substantial energy broadening. This is indeed observed in the experiment. Other effects, such as photoemission matrix elements could also contribute in suppressing the intensity of these bands, as they are clearly important and responsible for a strong variation in the observable bands throughout the extended BZ scheme. 

Finally, ARPES indicates that the crystals have a band gap of at least 1.5~eV. The strong charging below $T=$~160~K, however, suggests that the low temperature band gap that is significantly larger than that, more in line with the $\approx$~2.1~eV expected from the calculations.

\begin{acknowledgments}
This work was supported by VILLUM FONDEN via the Centre of Excellence for Dirac Materials (Grant No. 11744) and the Independent Research Fund Denmark  (Grant No. 1026-00089B). 
MR and MIK acknowledge the research program “Materials for the Quantum Age” (QuMat) for financial support. This program (registration number 024.005.006) is part of the Gravitation program financed by the Dutch Ministry of Education, Culture and Science (OCW).
MvS, SA and DP were supported the by the Computational Chemical Sciences program within the Office of Basic Energy Sciences, U.S. Department of Energy under Contract No. DE-AC36-08GO28308. FD was supported by the W{\"u}rzburg-Dresden Cluster of Excellence on Complexity and Topology in Quantum Matter ct.qmat (EXC 2147, ProjectID 390858490). MIK and SA are supported by the ERC Synergy Grant, project 854843 FASTCORR (Ultrafast dynamics of correlated
	electrons in solids). This research used resources of the National Energy Research Scientific Computing Center, a DOE Office of Science User Facility supported by the
Office of Science of the U.S. Department of Energy under Contract No. DE-AC02-05CH11231 using NERSC award BES-ERCAP0021783. SA and MIK acknowledge PRACE
	for awarding us access to Irene-Rome hosted by TGCC, France and Juwels Booster and Cluster, Germany.
\end{acknowledgments}

\section{Appendix}

\subsection{Additional Photoemission Data}

\begin{figure}
 \includegraphics[width=0.4\textwidth]{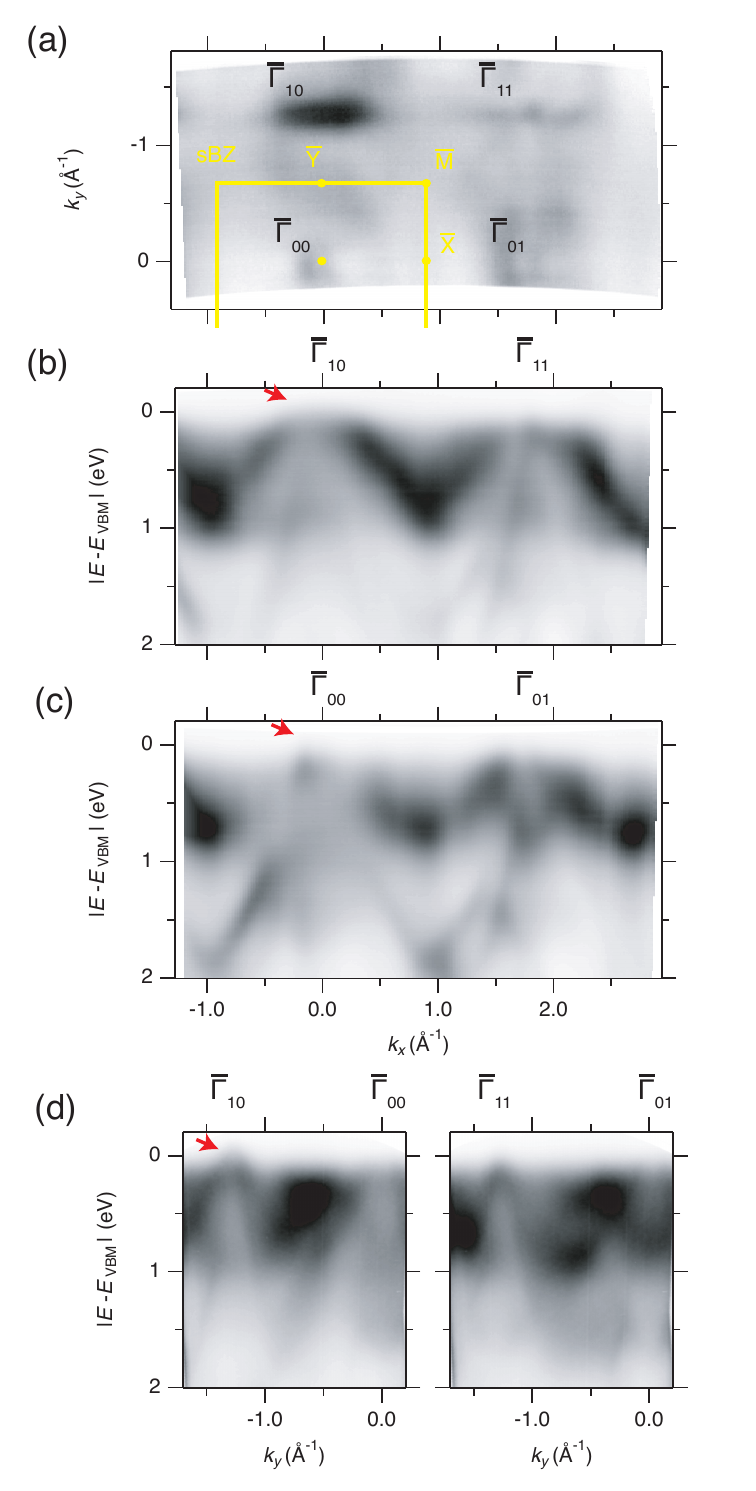}\\
  \caption{Same information as in Figs. 6 of the main text but with photoemission data collected at h$\nu$=74.6~eV.}
  \label{fig:5}
\end{figure}

\begin{figure}
 \includegraphics[width=0.5\textwidth]{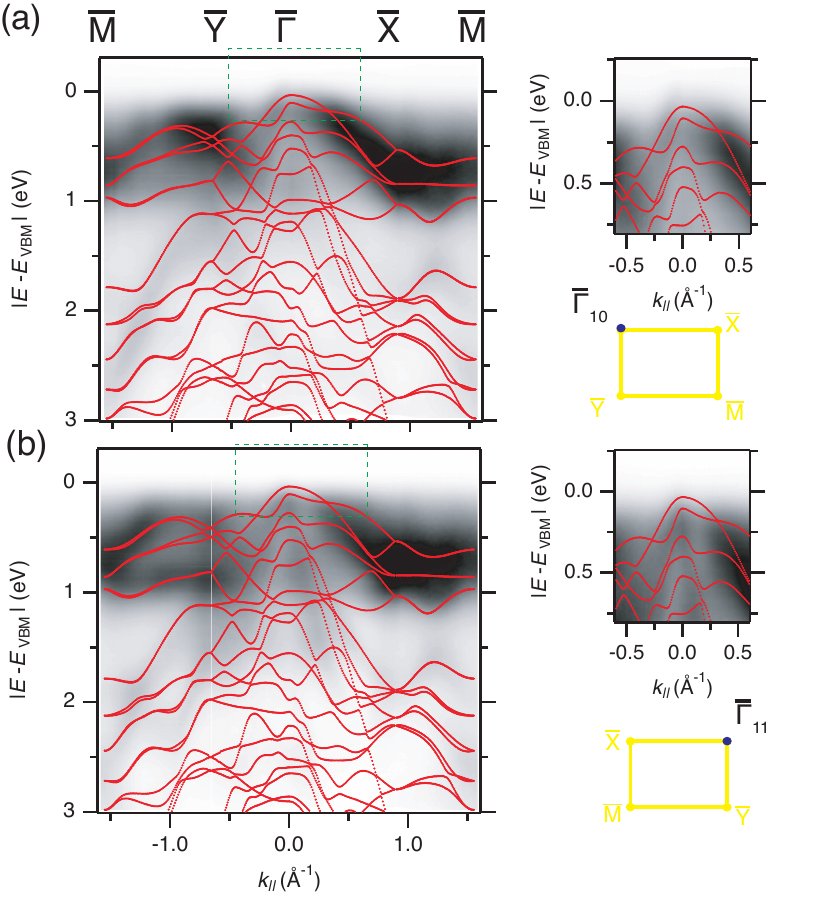}\\
  \caption{Same information as in Figs. 7 of the main text but with photoemission data collected at h$\nu$=74.6~eV.}
  \label{fig:6}
\end{figure}

Figs. \ref{fig:5} and \ref{fig:6} show ARPES data and a comparison between ARPES and the AFM-b QS$G\hat{W}$ calculation in the same way as Figs. 6 and 7 of the main text, respectively. The results are remarkably similar, including the situation around the VBM. This underlines the minor role of $k_z$ dispersion in this energy region.

\subsection{Effect of $k_z$ to quasi-particle band structure}

\begin{figure*}
  \includegraphics[width=0.666\textwidth]{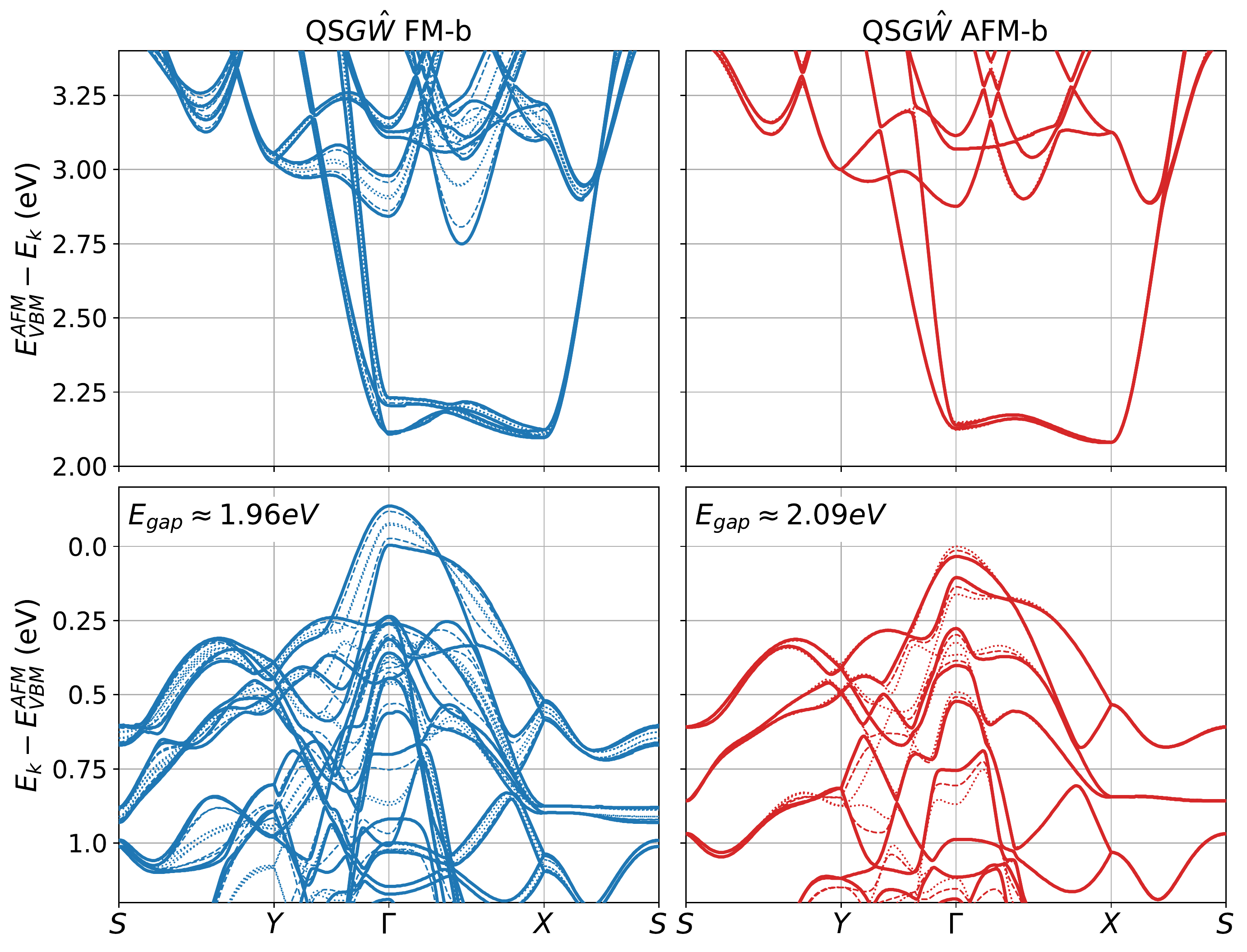}
  \caption{QS$G\hat{W}$ band structures in FM-b and AFM-b for different $k_z$. Solid lines: $k_z=0.0$, dashed lines: $k_z=0.25$, dotted lines: $k_z=0.5$ in units of $1/c$. \label{fig:theoryBSEKz}}
\end{figure*}

\begin{figure*}
  \includegraphics[width=0.99\textwidth]{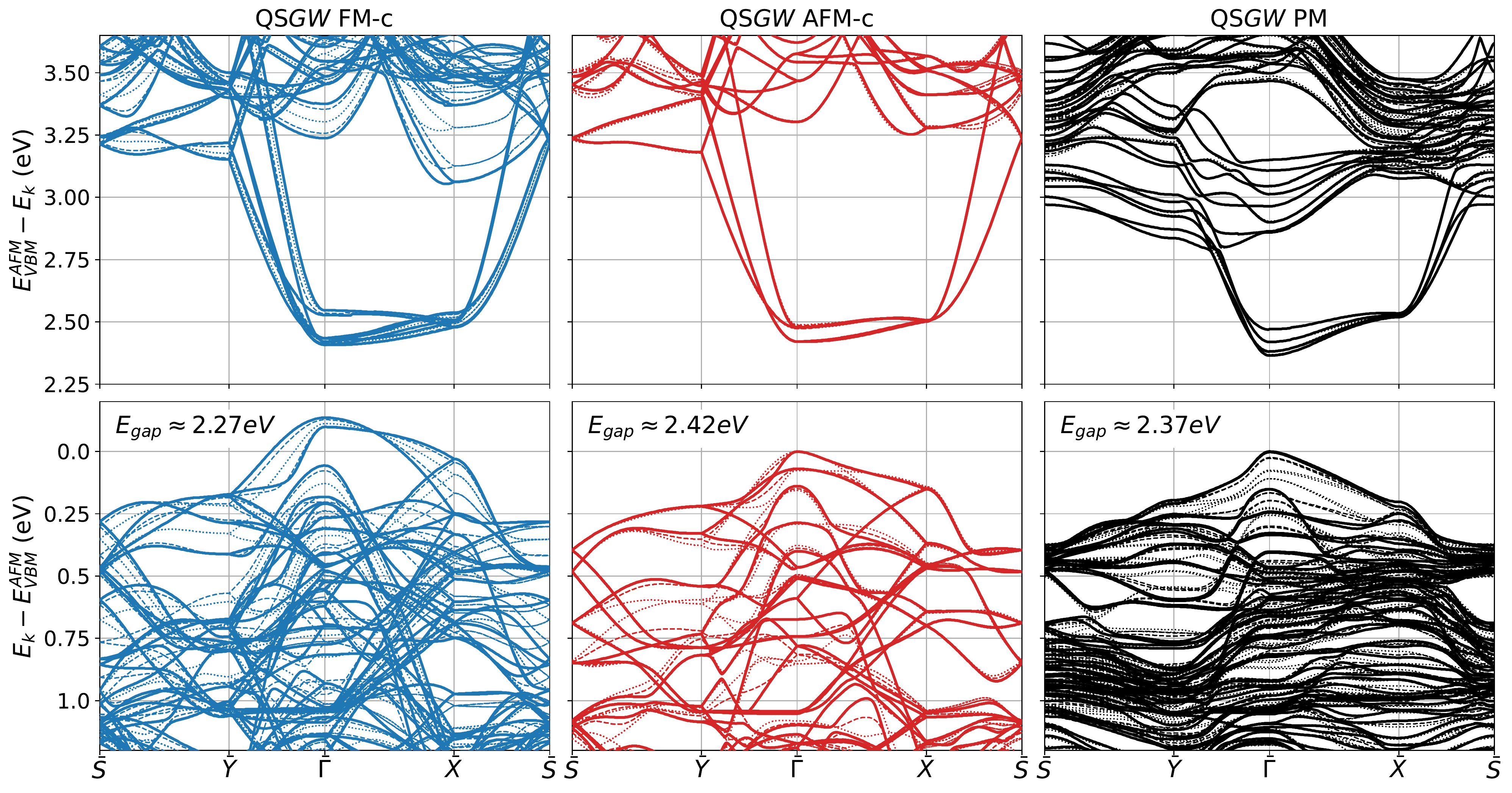}
  \caption{QS$GW$ band structures for supercells in FM-c and AFM-c together with the paramagnetic approximation for different $k_z$. Solid lines: $k_z=0.0$, dashed lines: $k_z=0.25$, dotted lines: $k_z=0.5$ in units of $1/c$. \label{fig:theoryGWSupKz}}
\end{figure*}

Figs.~\ref{fig:theoryBSEKz} and \ref{fig:theoryGWSupKz} depict the QS$G\hat{W}$ FM-b and AFM-b and QS$GW$ supercell band structures within the $k_\parallel$ plane for different $k_z$, respectively, including the conduction band states. While the AFM band structure show only minor $k_z$ dependencies, the FM and PM band structures are significantly dependent on the $k_z$ dispersion. In the corresponding energy and $k_\parallel$ ranges we can thus expect enhanced broadening effects within the PM ARPES data.


%\bibliographystyle{apsrev}
%\bibliography{phref,local,theory,temporary}

\begin{thebibliography}{42}
\expandafter\ifx\csname natexlab\endcsname\relax\def\natexlab#1{#1}\fi
\expandafter\ifx\csname bibnamefont\endcsname\relax
  \def\bibnamefont#1{#1}\fi
\expandafter\ifx\csname bibfnamefont\endcsname\relax
  \def\bibfnamefont#1{#1}\fi
\expandafter\ifx\csname citenamefont\endcsname\relax
  \def\citenamefont#1{#1}\fi
\expandafter\ifx\csname url\endcsname\relax
  \def\url#1{\texttt{#1}}\fi
\expandafter\ifx\csname urlprefix\endcsname\relax\def\urlprefix{URL }\fi
\providecommand{\bibinfo}[2]{#2}
\providecommand{\eprint}[2][]{\url{#2}}

\bibitem[{\citenamefont{Mermin and Wagner}(1966)}]{Mermin:1966uj}
\bibinfo{author}{\bibfnamefont{N.~D.} \bibnamefont{Mermin}} \bibnamefont{and}
  \bibinfo{author}{\bibfnamefont{H.}~\bibnamefont{Wagner}},
  \bibinfo{journal}{Phys. Rev. Lett.} \textbf{\bibinfo{volume}{17}},
  \bibinfo{pages}{1133} (\bibinfo{year}{1966}).

\bibitem[{\citenamefont{Huang et~al.}(2017)\citenamefont{Huang, Clark,
  Navarro-Moratalla, Klein, Cheng, Seyler, Zhong, Schmidgall, McGuire, Cobden
  et~al.}}]{Huang:2017aa}
\bibinfo{author}{\bibfnamefont{B.}~\bibnamefont{Huang}},
  \bibinfo{author}{\bibfnamefont{G.}~\bibnamefont{Clark}},
  \bibinfo{author}{\bibfnamefont{E.}~\bibnamefont{Navarro-Moratalla}},
  \bibinfo{author}{\bibfnamefont{D.~R.} \bibnamefont{Klein}},
  \bibinfo{author}{\bibfnamefont{R.}~\bibnamefont{Cheng}},
  \bibinfo{author}{\bibfnamefont{K.~L.} \bibnamefont{Seyler}},
  \bibinfo{author}{\bibfnamefont{D.}~\bibnamefont{Zhong}},
  \bibinfo{author}{\bibfnamefont{E.}~\bibnamefont{Schmidgall}},
  \bibinfo{author}{\bibfnamefont{M.~A.} \bibnamefont{McGuire}},
  \bibinfo{author}{\bibfnamefont{D.~H.} \bibnamefont{Cobden}},
  \bibnamefont{et~al.}, \bibinfo{journal}{Nature}
  \textbf{\bibinfo{volume}{546}}, \bibinfo{pages}{270} (\bibinfo{year}{2017}).

\bibitem[{\citenamefont{Gong et~al.}(2017)\citenamefont{Gong, Li, Li, Ji,
  Stern, Xia, Cao, Bao, Wang, Wang et~al.}}]{Gong:2017aa}
\bibinfo{author}{\bibfnamefont{C.}~\bibnamefont{Gong}},
  \bibinfo{author}{\bibfnamefont{L.}~\bibnamefont{Li}},
  \bibinfo{author}{\bibfnamefont{Z.}~\bibnamefont{Li}},
  \bibinfo{author}{\bibfnamefont{H.}~\bibnamefont{Ji}},
  \bibinfo{author}{\bibfnamefont{A.}~\bibnamefont{Stern}},
  \bibinfo{author}{\bibfnamefont{Y.}~\bibnamefont{Xia}},
  \bibinfo{author}{\bibfnamefont{T.}~\bibnamefont{Cao}},
  \bibinfo{author}{\bibfnamefont{W.}~\bibnamefont{Bao}},
  \bibinfo{author}{\bibfnamefont{C.}~\bibnamefont{Wang}},
  \bibinfo{author}{\bibfnamefont{Y.}~\bibnamefont{Wang}}, \bibnamefont{et~al.},
  \bibinfo{journal}{Nature} \textbf{\bibinfo{volume}{546}},
  \bibinfo{pages}{265} (\bibinfo{year}{2017}).

\bibitem[{\citenamefont{Deng et~al.}(2018)\citenamefont{Deng, Yu, Song, Zhang,
  Wang, Sun, Yi, Wu, Wu, Zhu et~al.}}]{Deng:2018aa}
\bibinfo{author}{\bibfnamefont{Y.}~\bibnamefont{Deng}},
  \bibinfo{author}{\bibfnamefont{Y.}~\bibnamefont{Yu}},
  \bibinfo{author}{\bibfnamefont{Y.}~\bibnamefont{Song}},
  \bibinfo{author}{\bibfnamefont{J.}~\bibnamefont{Zhang}},
  \bibinfo{author}{\bibfnamefont{N.~Z.} \bibnamefont{Wang}},
  \bibinfo{author}{\bibfnamefont{Z.}~\bibnamefont{Sun}},
  \bibinfo{author}{\bibfnamefont{Y.}~\bibnamefont{Yi}},
  \bibinfo{author}{\bibfnamefont{Y.~Z.} \bibnamefont{Wu}},
  \bibinfo{author}{\bibfnamefont{S.}~\bibnamefont{Wu}},
  \bibinfo{author}{\bibfnamefont{J.}~\bibnamefont{Zhu}}, \bibnamefont{et~al.},
  \bibinfo{journal}{Nature} \textbf{\bibinfo{volume}{563}}, \bibinfo{pages}{94}
  (\bibinfo{year}{2018}).

\bibitem[{\citenamefont{Telford et~al.}(2020)\citenamefont{Telford, Dismukes,
  Lee, Cheng, Wieteska, Bartholomew, Chen, Xu, Pasupathy, Zhu
  et~al.}}]{Telford:2020tc}
\bibinfo{author}{\bibfnamefont{E.~J.} \bibnamefont{Telford}},
  \bibinfo{author}{\bibfnamefont{A.~H.} \bibnamefont{Dismukes}},
  \bibinfo{author}{\bibfnamefont{K.}~\bibnamefont{Lee}},
  \bibinfo{author}{\bibfnamefont{M.}~\bibnamefont{Cheng}},
  \bibinfo{author}{\bibfnamefont{A.}~\bibnamefont{Wieteska}},
  \bibinfo{author}{\bibfnamefont{A.~K.} \bibnamefont{Bartholomew}},
  \bibinfo{author}{\bibfnamefont{Y.-S.} \bibnamefont{Chen}},
  \bibinfo{author}{\bibfnamefont{X.}~\bibnamefont{Xu}},
  \bibinfo{author}{\bibfnamefont{A.~N.} \bibnamefont{Pasupathy}},
  \bibinfo{author}{\bibfnamefont{X.}~\bibnamefont{Zhu}}, \bibnamefont{et~al.},
  \bibinfo{journal}{Advanced Materials} \textbf{\bibinfo{volume}{32}},
  \bibinfo{pages}{2003240} (\bibinfo{year}{2020}).

\bibitem[{\citenamefont{Klein et~al.}(2022{\natexlab{a}})\citenamefont{Klein,
  Pingault, Florian, Hei{\ss}enb{\"u}ttel, Steinhoff, Song, Torres, Dirnberger,
  Curtis, Deilmann et~al.}}]{Klein:2022ua}
\bibinfo{author}{\bibfnamefont{J.}~\bibnamefont{Klein}},
  \bibinfo{author}{\bibfnamefont{B.}~\bibnamefont{Pingault}},
  \bibinfo{author}{\bibfnamefont{M.}~\bibnamefont{Florian}},
  \bibinfo{author}{\bibfnamefont{M.-C.} \bibnamefont{Hei{\ss}enb{\"u}ttel}},
  \bibinfo{author}{\bibfnamefont{A.}~\bibnamefont{Steinhoff}},
  \bibinfo{author}{\bibfnamefont{Z.}~\bibnamefont{Song}},
  \bibinfo{author}{\bibfnamefont{K.}~\bibnamefont{Torres}},
  \bibinfo{author}{\bibfnamefont{F.}~\bibnamefont{Dirnberger}},
  \bibinfo{author}{\bibfnamefont{J.~B.} \bibnamefont{Curtis}},
  \bibinfo{author}{\bibfnamefont{T.}~\bibnamefont{Deilmann}},
  \bibnamefont{et~al.}, \emph{\bibinfo{title}{The bulk van der waals layered
  magnet crsbr is a quasi-1d quantum material}}
  (\bibinfo{year}{2022}{\natexlab{a}}).

\bibitem[{\citenamefont{G{\"o}ser et~al.}(1990)\citenamefont{G{\"o}ser, Paul,
  and Kahle}}]{Goser:1990aa}
\bibinfo{author}{\bibfnamefont{O.}~\bibnamefont{G{\"o}ser}},
  \bibinfo{author}{\bibfnamefont{W.}~\bibnamefont{Paul}}, \bibnamefont{and}
  \bibinfo{author}{\bibfnamefont{H.}~\bibnamefont{Kahle}},
  \bibinfo{journal}{Journal of Magnetism and Magnetic Materials}
  \textbf{\bibinfo{volume}{92}}, \bibinfo{pages}{129} (\bibinfo{year}{1990}).

\bibitem[{\citenamefont{Telford et~al.}(2022)\citenamefont{Telford, Dismukes,
  Dudley, Wiscons, Lee, Chica, Ziebel, Han, Yu, Shabani
  et~al.}}]{Telford:2022uy}
\bibinfo{author}{\bibfnamefont{E.~J.} \bibnamefont{Telford}},
  \bibinfo{author}{\bibfnamefont{A.~H.} \bibnamefont{Dismukes}},
  \bibinfo{author}{\bibfnamefont{R.~L.} \bibnamefont{Dudley}},
  \bibinfo{author}{\bibfnamefont{R.~A.} \bibnamefont{Wiscons}},
  \bibinfo{author}{\bibfnamefont{K.}~\bibnamefont{Lee}},
  \bibinfo{author}{\bibfnamefont{D.~G.} \bibnamefont{Chica}},
  \bibinfo{author}{\bibfnamefont{M.~E.} \bibnamefont{Ziebel}},
  \bibinfo{author}{\bibfnamefont{M.-G.} \bibnamefont{Han}},
  \bibinfo{author}{\bibfnamefont{J.}~\bibnamefont{Yu}},
  \bibinfo{author}{\bibfnamefont{S.}~\bibnamefont{Shabani}},
  \bibnamefont{et~al.}, \bibinfo{journal}{Nature Materials}
  \textbf{\bibinfo{volume}{21}}, \bibinfo{pages}{754} (\bibinfo{year}{2022}).

\bibitem[{\citenamefont{Wu et~al.}(2022)\citenamefont{Wu,
  Guti{\'{e}}rrez-Lezama, L{\'{o}}pez-Paz, Gibertini, Watanabe, Taniguchi, von
  Rohr, Ubrig, and Morpurgo}}]{Wu:2022}
\bibinfo{author}{\bibfnamefont{F.}~\bibnamefont{Wu}},
  \bibinfo{author}{\bibfnamefont{I.}~\bibnamefont{Guti{\'{e}}rrez-Lezama}},
  \bibinfo{author}{\bibfnamefont{S.~A.} \bibnamefont{L{\'{o}}pez-Paz}},
  \bibinfo{author}{\bibfnamefont{M.}~\bibnamefont{Gibertini}},
  \bibinfo{author}{\bibfnamefont{K.}~\bibnamefont{Watanabe}},
  \bibinfo{author}{\bibfnamefont{T.}~\bibnamefont{Taniguchi}},
  \bibinfo{author}{\bibfnamefont{F.~O.} \bibnamefont{von Rohr}},
  \bibinfo{author}{\bibfnamefont{N.}~\bibnamefont{Ubrig}}, \bibnamefont{and}
  \bibinfo{author}{\bibfnamefont{A.~F.} \bibnamefont{Morpurgo}},
  \bibinfo{journal}{Advanced Materials} \textbf{\bibinfo{volume}{34}},
  \bibinfo{pages}{2109759} (\bibinfo{year}{2022}).

\bibitem[{\citenamefont{Wilson et~al.}(2021)\citenamefont{Wilson, Lee, Cenker,
  Xie, Dismukes, Telford, Fonseca, Sivakumar, Dean, Cao
  et~al.}}]{Wilson:2021ty}
\bibinfo{author}{\bibfnamefont{N.~P.} \bibnamefont{Wilson}},
  \bibinfo{author}{\bibfnamefont{K.}~\bibnamefont{Lee}},
  \bibinfo{author}{\bibfnamefont{J.}~\bibnamefont{Cenker}},
  \bibinfo{author}{\bibfnamefont{K.}~\bibnamefont{Xie}},
  \bibinfo{author}{\bibfnamefont{A.~H.} \bibnamefont{Dismukes}},
  \bibinfo{author}{\bibfnamefont{E.~J.} \bibnamefont{Telford}},
  \bibinfo{author}{\bibfnamefont{J.}~\bibnamefont{Fonseca}},
  \bibinfo{author}{\bibfnamefont{S.}~\bibnamefont{Sivakumar}},
  \bibinfo{author}{\bibfnamefont{C.}~\bibnamefont{Dean}},
  \bibinfo{author}{\bibfnamefont{T.}~\bibnamefont{Cao}}, \bibnamefont{et~al.},
  \bibinfo{journal}{Nature Materials} \textbf{\bibinfo{volume}{20}},
  \bibinfo{pages}{1657} (\bibinfo{year}{2021}).

\bibitem[{\citenamefont{Torres et~al.}(2023)\citenamefont{Torres, Kuc, Maschio,
  Pham, Reidy, Dekanovsky, Sofer, Ross, and Klein}}]{Torres:2023}
\bibinfo{author}{\bibfnamefont{K.}~\bibnamefont{Torres}},
  \bibinfo{author}{\bibfnamefont{A.}~\bibnamefont{Kuc}},
  \bibinfo{author}{\bibfnamefont{L.}~\bibnamefont{Maschio}},
  \bibinfo{author}{\bibfnamefont{T.}~\bibnamefont{Pham}},
  \bibinfo{author}{\bibfnamefont{K.}~\bibnamefont{Reidy}},
  \bibinfo{author}{\bibfnamefont{L.}~\bibnamefont{Dekanovsky}},
  \bibinfo{author}{\bibfnamefont{Z.}~\bibnamefont{Sofer}},
  \bibinfo{author}{\bibfnamefont{F.~M.} \bibnamefont{Ross}}, \bibnamefont{and}
  \bibinfo{author}{\bibfnamefont{J.}~\bibnamefont{Klein}},
  \bibinfo{journal}{Advanced Functional Materials} p. \bibinfo{pages}{2211366}
  (\bibinfo{year}{2023}).

\bibitem[{\citenamefont{Pawbake et~al.}(2023)\citenamefont{Pawbake, Pelini,
  Wilson, Mosina, Sofer, Heid, and Faugeras}}]{Pawbake.2023}
\bibinfo{author}{\bibfnamefont{A.}~\bibnamefont{Pawbake}},
  \bibinfo{author}{\bibfnamefont{T.}~\bibnamefont{Pelini}},
  \bibinfo{author}{\bibfnamefont{N.~P.} \bibnamefont{Wilson}},
  \bibinfo{author}{\bibfnamefont{K.}~\bibnamefont{Mosina}},
  \bibinfo{author}{\bibfnamefont{Z.}~\bibnamefont{Sofer}},
  \bibinfo{author}{\bibfnamefont{R.}~\bibnamefont{Heid}}, \bibnamefont{and}
  \bibinfo{author}{\bibfnamefont{C.}~\bibnamefont{Faugeras}},
  \bibinfo{journal}{Physical Review B} \textbf{\bibinfo{volume}{107}}
  (\bibinfo{year}{2023}).

\bibitem[{\citenamefont{Bae et~al.}(2022)\citenamefont{Bae, Wang, Scheie, Xu,
  Chica, Diederich, Cenker, Ziebel, Bai, Ren et~al.}}]{Bae:2022}
\bibinfo{author}{\bibfnamefont{Y.~J.} \bibnamefont{Bae}},
  \bibinfo{author}{\bibfnamefont{J.}~\bibnamefont{Wang}},
  \bibinfo{author}{\bibfnamefont{A.}~\bibnamefont{Scheie}},
  \bibinfo{author}{\bibfnamefont{J.}~\bibnamefont{Xu}},
  \bibinfo{author}{\bibfnamefont{D.~G.} \bibnamefont{Chica}},
  \bibinfo{author}{\bibfnamefont{G.~M.} \bibnamefont{Diederich}},
  \bibinfo{author}{\bibfnamefont{J.}~\bibnamefont{Cenker}},
  \bibinfo{author}{\bibfnamefont{M.~E.} \bibnamefont{Ziebel}},
  \bibinfo{author}{\bibfnamefont{Y.}~\bibnamefont{Bai}},
  \bibinfo{author}{\bibfnamefont{H.}~\bibnamefont{Ren}}, \bibnamefont{et~al.},
  \bibinfo{journal}{Nature} \textbf{\bibinfo{volume}{609}},
  \bibinfo{pages}{282} (\bibinfo{year}{2022}).

\bibitem[{\citenamefont{Guo et~al.}(2018)\citenamefont{Guo, Zhang, Yuan, Wang,
  and Wang}}]{Guo:2018wt}
\bibinfo{author}{\bibfnamefont{Y.}~\bibnamefont{Guo}},
  \bibinfo{author}{\bibfnamefont{Y.}~\bibnamefont{Zhang}},
  \bibinfo{author}{\bibfnamefont{S.}~\bibnamefont{Yuan}},
  \bibinfo{author}{\bibfnamefont{B.}~\bibnamefont{Wang}}, \bibnamefont{and}
  \bibinfo{author}{\bibfnamefont{J.}~\bibnamefont{Wang}},
  \bibinfo{journal}{Nanoscale} \textbf{\bibinfo{volume}{10}},
  \bibinfo{pages}{18036} (\bibinfo{year}{2018}).

\bibitem[{\citenamefont{Jiang et~al.}(2018)\citenamefont{Jiang, Wang, Xing,
  Jiang, and Zhao}}]{Jiang:2018ug}
\bibinfo{author}{\bibfnamefont{Z.}~\bibnamefont{Jiang}},
  \bibinfo{author}{\bibfnamefont{P.}~\bibnamefont{Wang}},
  \bibinfo{author}{\bibfnamefont{J.}~\bibnamefont{Xing}},
  \bibinfo{author}{\bibfnamefont{X.}~\bibnamefont{Jiang}}, \bibnamefont{and}
  \bibinfo{author}{\bibfnamefont{J.}~\bibnamefont{Zhao}},
  \bibinfo{journal}{{ACS} Applied Materials and Interfaces}
  \textbf{\bibinfo{volume}{10}}, \bibinfo{pages}{39032} (\bibinfo{year}{2018}).

\bibitem[{\citenamefont{Wang et~al.}(2019)\citenamefont{Wang, Zhou, Zhou, Tong,
  Lu, and Ji}}]{Wang:2019wf}
\bibinfo{author}{\bibfnamefont{C.}~\bibnamefont{Wang}},
  \bibinfo{author}{\bibfnamefont{X.}~\bibnamefont{Zhou}},
  \bibinfo{author}{\bibfnamefont{L.}~\bibnamefont{Zhou}},
  \bibinfo{author}{\bibfnamefont{N.-H.} \bibnamefont{Tong}},
  \bibinfo{author}{\bibfnamefont{Z.-Y.} \bibnamefont{Lu}}, \bibnamefont{and}
  \bibinfo{author}{\bibfnamefont{W.}~\bibnamefont{Ji}},
  \bibinfo{journal}{Science Bulletin} \textbf{\bibinfo{volume}{64}},
  \bibinfo{pages}{293} (\bibinfo{year}{2019}).

\bibitem[{\citenamefont{Wang et~al.}(2020)\citenamefont{Wang, Qi, and
  Qian}}]{Wang:2020wt}
\bibinfo{author}{\bibfnamefont{H.}~\bibnamefont{Wang}},
  \bibinfo{author}{\bibfnamefont{J.}~\bibnamefont{Qi}}, \bibnamefont{and}
  \bibinfo{author}{\bibfnamefont{X.}~\bibnamefont{Qian}},
  \bibinfo{journal}{Applied Physics Letters} \textbf{\bibinfo{volume}{117}},
  \bibinfo{pages}{083102} (\bibinfo{year}{2020}).

\bibitem[{\citenamefont{Xu et~al.}(2022)\citenamefont{Xu, Wang, Chang, Chen,
  Guan, and Tao}}]{Xu:2022uo}
\bibinfo{author}{\bibfnamefont{X.}~\bibnamefont{Xu}},
  \bibinfo{author}{\bibfnamefont{X.}~\bibnamefont{Wang}},
  \bibinfo{author}{\bibfnamefont{P.}~\bibnamefont{Chang}},
  \bibinfo{author}{\bibfnamefont{X.}~\bibnamefont{Chen}},
  \bibinfo{author}{\bibfnamefont{L.}~\bibnamefont{Guan}}, \bibnamefont{and}
  \bibinfo{author}{\bibfnamefont{J.}~\bibnamefont{Tao}}, \bibinfo{journal}{The
  Journal of Physical Chemistry C} \textbf{\bibinfo{volume}{126}},
  \bibinfo{pages}{10574} (\bibinfo{year}{2022}).

\bibitem[{\citenamefont{Klein et~al.}(2022{\natexlab{b}})\citenamefont{Klein,
  Pham, Thomsen, Curtis, Denneulin, Lorke, Florian, Steinhoff, Wiscons, Luxa
  et~al.}}]{Klein:2022}
\bibinfo{author}{\bibfnamefont{J.}~\bibnamefont{Klein}},
  \bibinfo{author}{\bibfnamefont{T.}~\bibnamefont{Pham}},
  \bibinfo{author}{\bibfnamefont{J.~D.} \bibnamefont{Thomsen}},
  \bibinfo{author}{\bibfnamefont{J.~B.} \bibnamefont{Curtis}},
  \bibinfo{author}{\bibfnamefont{T.}~\bibnamefont{Denneulin}},
  \bibinfo{author}{\bibfnamefont{M.}~\bibnamefont{Lorke}},
  \bibinfo{author}{\bibfnamefont{M.}~\bibnamefont{Florian}},
  \bibinfo{author}{\bibfnamefont{A.}~\bibnamefont{Steinhoff}},
  \bibinfo{author}{\bibfnamefont{R.~A.} \bibnamefont{Wiscons}},
  \bibinfo{author}{\bibfnamefont{J.}~\bibnamefont{Luxa}}, \bibnamefont{et~al.},
  \bibinfo{journal}{Nature Communications} \textbf{\bibinfo{volume}{13}}
  (\bibinfo{year}{2022}{\natexlab{b}}).

\bibitem[{\citenamefont{van Schilfgaarde et~al.}(2006)\citenamefont{van
  Schilfgaarde, Kotani, and Faleev}}]{qsgw}
\bibinfo{author}{\bibfnamefont{M.}~\bibnamefont{van Schilfgaarde}},
  \bibinfo{author}{\bibfnamefont{T.}~\bibnamefont{Kotani}}, \bibnamefont{and}
  \bibinfo{author}{\bibfnamefont{S.}~\bibnamefont{Faleev}},
  \bibinfo{journal}{Physical review letters} \textbf{\bibinfo{volume}{96}},
  \bibinfo{pages}{226402} (\bibinfo{year}{2006}).

\bibitem[{\citenamefont{Ismail-Beigi}(2017)}]{variational}
\bibinfo{author}{\bibfnamefont{S.}~\bibnamefont{Ismail-Beigi}},
  \bibinfo{journal}{Journal of Physics: Condensed Matter}
  \textbf{\bibinfo{volume}{29}}, \bibinfo{pages}{385501}
  (\bibinfo{year}{2017}).

\bibitem[{\citenamefont{Cunningham et~al.}(2023)\citenamefont{Cunningham,
  Gruening, Pashov, and van Schilfgaarde}}]{Cunningham2023}
\bibinfo{author}{\bibfnamefont{B.}~\bibnamefont{Cunningham}},
  \bibinfo{author}{\bibfnamefont{M.}~\bibnamefont{Gruening}},
  \bibinfo{author}{\bibfnamefont{D.}~\bibnamefont{Pashov}}, \bibnamefont{and}
  \bibinfo{author}{\bibfnamefont{M.}~\bibnamefont{van Schilfgaarde}}
  (\bibinfo{year}{2023}), \bibinfo{note}{preprint arXiv 2302.06325}.

\bibitem[{\citenamefont{Hirata and Head-Gordon}(1999)}]{hirata1999}
\bibinfo{author}{\bibfnamefont{S.}~\bibnamefont{Hirata}} \bibnamefont{and}
  \bibinfo{author}{\bibfnamefont{M.}~\bibnamefont{Head-Gordon}},
  \bibinfo{journal}{Chemical Physics Letters} \textbf{\bibinfo{volume}{314}},
  \bibinfo{pages}{291} (\bibinfo{year}{1999}).

\bibitem[{\citenamefont{Acharya
  et~al.}(2021{\natexlab{a}})\citenamefont{Acharya, Pashov, Rudenko,
  R{\"o}sner, van Schilfgaarde, and Katsnelson}}]{acharya2021importance}
\bibinfo{author}{\bibfnamefont{S.}~\bibnamefont{Acharya}},
  \bibinfo{author}{\bibfnamefont{D.}~\bibnamefont{Pashov}},
  \bibinfo{author}{\bibfnamefont{A.~N.} \bibnamefont{Rudenko}},
  \bibinfo{author}{\bibfnamefont{M.}~\bibnamefont{R{\"o}sner}},
  \bibinfo{author}{\bibfnamefont{M.}~\bibnamefont{van Schilfgaarde}},
  \bibnamefont{and} \bibinfo{author}{\bibfnamefont{M.~I.}
  \bibnamefont{Katsnelson}}, \bibinfo{journal}{npj Computational Materials}
  \textbf{\bibinfo{volume}{7}}, \bibinfo{pages}{1}
  (\bibinfo{year}{2021}{\natexlab{a}}).

\bibitem[{\citenamefont{Göser et~al.}(1990)\citenamefont{Göser, Paul, and
  Kahle}}]{goser_magnetic_1990}
\bibinfo{author}{\bibfnamefont{O.}~\bibnamefont{Göser}},
  \bibinfo{author}{\bibfnamefont{W.}~\bibnamefont{Paul}}, \bibnamefont{and}
  \bibinfo{author}{\bibfnamefont{H.~G.} \bibnamefont{Kahle}},
  \textbf{\bibinfo{volume}{92}}, \bibinfo{pages}{129} (\bibinfo{year}{1990}).

\bibitem[{\citenamefont{López-Paz et~al.}(2022)\citenamefont{López-Paz,
  Guguchia, Pomjakushin, Witteveen, Cervellino, Luetkens, Casati, Morpurgo, and
  von Rohr}}]{lopez-paz_dynamic_2022}
\bibinfo{author}{\bibfnamefont{S.~A.} \bibnamefont{López-Paz}},
  \bibinfo{author}{\bibfnamefont{Z.}~\bibnamefont{Guguchia}},
  \bibinfo{author}{\bibfnamefont{V.~Y.} \bibnamefont{Pomjakushin}},
  \bibinfo{author}{\bibfnamefont{C.}~\bibnamefont{Witteveen}},
  \bibinfo{author}{\bibfnamefont{A.}~\bibnamefont{Cervellino}},
  \bibinfo{author}{\bibfnamefont{H.}~\bibnamefont{Luetkens}},
  \bibinfo{author}{\bibfnamefont{N.}~\bibnamefont{Casati}},
  \bibinfo{author}{\bibfnamefont{A.~F.} \bibnamefont{Morpurgo}},
  \bibnamefont{and} \bibinfo{author}{\bibfnamefont{F.~O.} \bibnamefont{von
  Rohr}}, \textbf{\bibinfo{volume}{13}}, \bibinfo{pages}{4745}
  (\bibinfo{year}{2022}).

\bibitem[{\citenamefont{Kresse and Furthmüller}(1996)}]{Kresse1}
\bibinfo{author}{\bibfnamefont{G.}~\bibnamefont{Kresse}} \bibnamefont{and}
  \bibinfo{author}{\bibfnamefont{J.}~\bibnamefont{Furthmüller}},
  \bibinfo{journal}{Comp. Mat. Sci.} \textbf{\bibinfo{volume}{6}},
  \bibinfo{pages}{15} (\bibinfo{year}{1996}).

\bibitem[{\citenamefont{Kresse and Furthm\"uller}(1996)}]{Kresse2}
\bibinfo{author}{\bibfnamefont{G.}~\bibnamefont{Kresse}} \bibnamefont{and}
  \bibinfo{author}{\bibfnamefont{J.}~\bibnamefont{Furthm\"uller}},
  \bibinfo{journal}{Phys. Rev. B} \textbf{\bibinfo{volume}{54}},
  \bibinfo{pages}{11169} (\bibinfo{year}{1996}).

\bibitem[{\citenamefont{Perdew et~al.}(1996)\citenamefont{Perdew, Burke, and
  Ernzerhof}}]{pbe}
\bibinfo{author}{\bibfnamefont{J.~P.} \bibnamefont{Perdew}},
  \bibinfo{author}{\bibfnamefont{K.}~\bibnamefont{Burke}}, \bibnamefont{and}
  \bibinfo{author}{\bibfnamefont{M.}~\bibnamefont{Ernzerhof}},
  \bibinfo{journal}{Phys. Rev. Lett.} \textbf{\bibinfo{volume}{77}},
  \bibinfo{pages}{3865} (\bibinfo{year}{1996}).

\bibitem[{\citenamefont{Lee et~al.}(2021{\natexlab{a}})\citenamefont{Lee,
  Dismukes, Telford, Wiscons, Wang, Xu, Nuckolls, Dean, Roy, and
  Zhu}}]{lee_magnetic_2021}
\bibinfo{author}{\bibfnamefont{K.}~\bibnamefont{Lee}},
  \bibinfo{author}{\bibfnamefont{A.~H.} \bibnamefont{Dismukes}},
  \bibinfo{author}{\bibfnamefont{E.~J.} \bibnamefont{Telford}},
  \bibinfo{author}{\bibfnamefont{R.~A.} \bibnamefont{Wiscons}},
  \bibinfo{author}{\bibfnamefont{J.}~\bibnamefont{Wang}},
  \bibinfo{author}{\bibfnamefont{X.}~\bibnamefont{Xu}},
  \bibinfo{author}{\bibfnamefont{C.}~\bibnamefont{Nuckolls}},
  \bibinfo{author}{\bibfnamefont{C.~R.} \bibnamefont{Dean}},
  \bibinfo{author}{\bibfnamefont{X.}~\bibnamefont{Roy}}, \bibnamefont{and}
  \bibinfo{author}{\bibfnamefont{X.}~\bibnamefont{Zhu}},
  \textbf{\bibinfo{volume}{21}}, \bibinfo{pages}{3511}
  (\bibinfo{year}{2021}{\natexlab{a}}).

\bibitem[{\citenamefont{Zunger et~al.}(1990)\citenamefont{Zunger, Wei,
  Ferreira, and Bernard}}]{zunger}
\bibinfo{author}{\bibfnamefont{A.}~\bibnamefont{Zunger}},
  \bibinfo{author}{\bibfnamefont{S.-H.} \bibnamefont{Wei}},
  \bibinfo{author}{\bibfnamefont{L.~G.} \bibnamefont{Ferreira}},
  \bibnamefont{and} \bibinfo{author}{\bibfnamefont{J.~E.}
  \bibnamefont{Bernard}}, \bibinfo{journal}{Phys. Rev. Lett.}
  \textbf{\bibinfo{volume}{65}}, \bibinfo{pages}{353} (\bibinfo{year}{1990}).

\bibitem[{\citenamefont{Hoffmann et~al.}(2004)\citenamefont{Hoffmann,
  S{\o}ndergaard, Schultz, Li, and Hofmann}}]{Hoffmann:2004aa}
\bibinfo{author}{\bibfnamefont{S.~V.} \bibnamefont{Hoffmann}},
  \bibinfo{author}{\bibfnamefont{C.}~\bibnamefont{S{\o}ndergaard}},
  \bibinfo{author}{\bibfnamefont{C.}~\bibnamefont{Schultz}},
  \bibinfo{author}{\bibfnamefont{Z.}~\bibnamefont{Li}}, \bibnamefont{and}
  \bibinfo{author}{\bibfnamefont{P.}~\bibnamefont{Hofmann}},
  \bibinfo{journal}{Nuclear Instruments and Methods in Physics Research, A}
  \textbf{\bibinfo{volume}{523}}, \bibinfo{pages}{441} (\bibinfo{year}{2004}).

\bibitem[{\citenamefont{Acharya
  et~al.}(2021{\natexlab{b}})\citenamefont{Acharya, Pashov, Cunningham,
  Rudenko, R{\"o}sner, Gr{\"u}ning, van Schilfgaarde, and
  Katsnelson}}]{acharya2021electronic}
\bibinfo{author}{\bibfnamefont{S.}~\bibnamefont{Acharya}},
  \bibinfo{author}{\bibfnamefont{D.}~\bibnamefont{Pashov}},
  \bibinfo{author}{\bibfnamefont{B.}~\bibnamefont{Cunningham}},
  \bibinfo{author}{\bibfnamefont{A.~N.} \bibnamefont{Rudenko}},
  \bibinfo{author}{\bibfnamefont{M.}~\bibnamefont{R{\"o}sner}},
  \bibinfo{author}{\bibfnamefont{M.}~\bibnamefont{Gr{\"u}ning}},
  \bibinfo{author}{\bibfnamefont{M.}~\bibnamefont{van Schilfgaarde}},
  \bibnamefont{and} \bibinfo{author}{\bibfnamefont{M.~I.}
  \bibnamefont{Katsnelson}}, \bibinfo{journal}{Physical Review B}
  \textbf{\bibinfo{volume}{104}}, \bibinfo{pages}{155109}
  (\bibinfo{year}{2021}{\natexlab{b}}).

\bibitem[{\citenamefont{Acharya et~al.}(2022)\citenamefont{Acharya, Pashov,
  Rudenko, R{\"o}sner, Schilfgaarde, and Katsnelson}}]{acharya2022real}
\bibinfo{author}{\bibfnamefont{S.}~\bibnamefont{Acharya}},
  \bibinfo{author}{\bibfnamefont{D.}~\bibnamefont{Pashov}},
  \bibinfo{author}{\bibfnamefont{A.~N.} \bibnamefont{Rudenko}},
  \bibinfo{author}{\bibfnamefont{M.}~\bibnamefont{R{\"o}sner}},
  \bibinfo{author}{\bibfnamefont{M.~v.} \bibnamefont{Schilfgaarde}},
  \bibnamefont{and} \bibinfo{author}{\bibfnamefont{M.~I.}
  \bibnamefont{Katsnelson}}, \bibinfo{journal}{npj 2D Materials and
  Applications} \textbf{\bibinfo{volume}{6}}, \bibinfo{pages}{1}
  (\bibinfo{year}{2022}).

\bibitem[{\citenamefont{Bruneval et~al.}(2006)\citenamefont{Bruneval, Vast,
  Reining, Izquierdo, Sirotti, and Barrett}}]{Bruneval06b}
\bibinfo{author}{\bibfnamefont{F.}~\bibnamefont{Bruneval}},
  \bibinfo{author}{\bibfnamefont{N.}~\bibnamefont{Vast}},
  \bibinfo{author}{\bibfnamefont{L.}~\bibnamefont{Reining}},
  \bibinfo{author}{\bibfnamefont{M.}~\bibnamefont{Izquierdo}},
  \bibinfo{author}{\bibfnamefont{F.}~\bibnamefont{Sirotti}}, \bibnamefont{and}
  \bibinfo{author}{\bibfnamefont{N.}~\bibnamefont{Barrett}},
  \bibinfo{journal}{Phys. Rev. Lett.} \textbf{\bibinfo{volume}{97}},
  \bibinfo{pages}{267601} (\bibinfo{year}{2006}).

\bibitem[{\citenamefont{Vidal et~al.}(2010)\citenamefont{Vidal, Botti, Olsson,
  Guillemoles, and Reining}}]{Vidal10}
\bibinfo{author}{\bibfnamefont{J.}~\bibnamefont{Vidal}},
  \bibinfo{author}{\bibfnamefont{S.}~\bibnamefont{Botti}},
  \bibinfo{author}{\bibfnamefont{P.}~\bibnamefont{Olsson}},
  \bibinfo{author}{\bibfnamefont{J.-F.} \bibnamefont{Guillemoles}},
  \bibnamefont{and} \bibinfo{author}{\bibfnamefont{L.}~\bibnamefont{Reining}},
  \bibinfo{journal}{Phys. Rev. Lett.} \textbf{\bibinfo{volume}{104}},
  \bibinfo{pages}{056401} (\bibinfo{year}{2010}).

\bibitem[{\citenamefont{Lee et~al.}(2021{\natexlab{b}})\citenamefont{Lee,
  Dismukes, Telford, Wiscons, Wang, Xu, Nuckolls, Dean, Roy, and
  Zhu}}]{Lee:2021to}
\bibinfo{author}{\bibfnamefont{K.}~\bibnamefont{Lee}},
  \bibinfo{author}{\bibfnamefont{A.~H.} \bibnamefont{Dismukes}},
  \bibinfo{author}{\bibfnamefont{E.~J.} \bibnamefont{Telford}},
  \bibinfo{author}{\bibfnamefont{R.~A.} \bibnamefont{Wiscons}},
  \bibinfo{author}{\bibfnamefont{J.}~\bibnamefont{Wang}},
  \bibinfo{author}{\bibfnamefont{X.}~\bibnamefont{Xu}},
  \bibinfo{author}{\bibfnamefont{C.}~\bibnamefont{Nuckolls}},
  \bibinfo{author}{\bibfnamefont{C.~R.} \bibnamefont{Dean}},
  \bibinfo{author}{\bibfnamefont{X.}~\bibnamefont{Roy}}, \bibnamefont{and}
  \bibinfo{author}{\bibfnamefont{X.}~\bibnamefont{Zhu}}, \bibinfo{journal}{Nano
  Letters} \textbf{\bibinfo{volume}{21}}, \bibinfo{pages}{3511}
  (\bibinfo{year}{2021}{\natexlab{b}}).

\bibitem[{\citenamefont{Nakayama et~al.}(2020)\citenamefont{Nakayama, Kera, and
  Ueno}}]{Nakayama2020}
\bibinfo{author}{\bibfnamefont{Y.}~\bibnamefont{Nakayama}},
  \bibinfo{author}{\bibfnamefont{S.}~\bibnamefont{Kera}}, \bibnamefont{and}
  \bibinfo{author}{\bibfnamefont{N.}~\bibnamefont{Ueno}},
  \bibinfo{journal}{Journal of Materials Chemistry C}
  \textbf{\bibinfo{volume}{8}}, \bibinfo{pages}{9090} (\bibinfo{year}{2020}).

\bibitem[{\citenamefont{Shirley et~al.}(1995)\citenamefont{Shirley, Terminello,
  Santoni, and Himpsel}}]{Shirley:1995aa}
\bibinfo{author}{\bibfnamefont{E.~L.} \bibnamefont{Shirley}},
  \bibinfo{author}{\bibfnamefont{L.~J.} \bibnamefont{Terminello}},
  \bibinfo{author}{\bibfnamefont{A.}~\bibnamefont{Santoni}}, \bibnamefont{and}
  \bibinfo{author}{\bibfnamefont{F.~J.} \bibnamefont{Himpsel}},
  \bibinfo{journal}{Phys. Rev. B} \textbf{\bibinfo{volume}{51}},
  \bibinfo{pages}{13614} (\bibinfo{year}{1995}).

\bibitem[{\citenamefont{Lizzit et~al.}(2010)\citenamefont{Lizzit, Zampieri,
  Petaccia, Larciprete, Lacovig, Rienks, Bihlmayer, Baraldi, and
  Hofmann}}]{Lizzit:2010aa}
\bibinfo{author}{\bibfnamefont{S.}~\bibnamefont{Lizzit}},
  \bibinfo{author}{\bibfnamefont{G.}~\bibnamefont{Zampieri}},
  \bibinfo{author}{\bibfnamefont{L.}~\bibnamefont{Petaccia}},
  \bibinfo{author}{\bibfnamefont{R.}~\bibnamefont{Larciprete}},
  \bibinfo{author}{\bibfnamefont{P.}~\bibnamefont{Lacovig}},
  \bibinfo{author}{\bibfnamefont{E.~D.~L.} \bibnamefont{Rienks}},
  \bibinfo{author}{\bibfnamefont{G.}~\bibnamefont{Bihlmayer}},
  \bibinfo{author}{\bibfnamefont{A.}~\bibnamefont{Baraldi}}, \bibnamefont{and}
  \bibinfo{author}{\bibfnamefont{P.}~\bibnamefont{Hofmann}},
  \bibinfo{journal}{Nature Physics} \textbf{\bibinfo{volume}{6}},
  \bibinfo{pages}{345} (\bibinfo{year}{2010}).

\bibitem[{\citenamefont{Mazzola et~al.}(2013)\citenamefont{Mazzola, Wells,
  Yakimova, Ulstrup, Miwa, Balog, Bianchi, Leandersson, Adell, Hofmann
  et~al.}}]{Mazzola:2013aa}
\bibinfo{author}{\bibfnamefont{F.}~\bibnamefont{Mazzola}},
  \bibinfo{author}{\bibfnamefont{J.~W.} \bibnamefont{Wells}},
  \bibinfo{author}{\bibfnamefont{R.}~\bibnamefont{Yakimova}},
  \bibinfo{author}{\bibfnamefont{S.}~\bibnamefont{Ulstrup}},
  \bibinfo{author}{\bibfnamefont{J.~A.} \bibnamefont{Miwa}},
  \bibinfo{author}{\bibfnamefont{R.}~\bibnamefont{Balog}},
  \bibinfo{author}{\bibfnamefont{M.}~\bibnamefont{Bianchi}},
  \bibinfo{author}{\bibfnamefont{M.}~\bibnamefont{Leandersson}},
  \bibinfo{author}{\bibfnamefont{J.}~\bibnamefont{Adell}},
  \bibinfo{author}{\bibfnamefont{P.}~\bibnamefont{Hofmann}},
  \bibnamefont{et~al.}, \bibinfo{journal}{Phys. Rev. Lett.}
  \textbf{\bibinfo{volume}{111}}, \bibinfo{pages}{216806}
  (\bibinfo{year}{2013}).

\bibitem[{\citenamefont{Mazzola et~al.}(2017)\citenamefont{Mazzola,
  Frederiksen, Balasubramanian, Hofmann, Hellsing, and Wells}}]{Mazzola:2017aa}
\bibinfo{author}{\bibfnamefont{F.}~\bibnamefont{Mazzola}},
  \bibinfo{author}{\bibfnamefont{T.}~\bibnamefont{Frederiksen}},
  \bibinfo{author}{\bibfnamefont{T.}~\bibnamefont{Balasubramanian}},
  \bibinfo{author}{\bibfnamefont{P.}~\bibnamefont{Hofmann}},
  \bibinfo{author}{\bibfnamefont{B.}~\bibnamefont{Hellsing}}, \bibnamefont{and}
  \bibinfo{author}{\bibfnamefont{J.~W.} \bibnamefont{Wells}},
  \bibinfo{journal}{Phys. Rev. B} \textbf{\bibinfo{volume}{95}},
  \bibinfo{pages}{075430} (\bibinfo{year}{2017}).

\end{thebibliography}

\end{document}